\documentclass[journal]{IEEEtran}

\usepackage{color, colortbl}
\definecolor{Gray}{gray}{0.9}
\definecolor{LightCyan}{rgb}{0.88,1,1}
\definecolor{LightRed}{rgb}{1,0.88,0.88}
\definecolor{LightBlue}{rgb}{0.12,0.56,0.8}
\usepackage[table]{xcolor}

\usepackage[first=0,last=9]{lcg}

\usepackage{xfrac} 

\usepackage{cite}

\usepackage{graphicx}
\ifCLASSINFOpdf
\else
\fi
\usepackage{amsmath}
\usepackage[linesnumbered,ruled,vlined]{algorithm2e}
\SetKwProg{Fn}{Function}{}{}

\SetCommentSty{mycommfont}

\makeatletter
\newcommand{\removelatexerror}{\let\@latex@error\@gobble}
\makeatother

\usepackage{array}

\ifCLASSOPTIONcompsoc
  \usepackage[caption=false,font=normalsize,labelfont=sf,textfont=sf]{subfig}
\else
  \usepackage[caption=false,font=footnotesize]{subfig}
\fi

\usepackage{dblfloatfix}    

\usepackage{url}

\begin{document}
%
\title{C2TCP: A Flexible Cellular TCP to Meet Stringent Delay Requirements}
%
%
%
\author{Soheil~Abbasloo, Yang~Xu,~\IEEEmembership{Member,~IEEE,} H.~Jonathan~Chao,~\IEEEmembership{Fellow,~IEEE}%
}
\maketitle

\begin{abstract}
Since current widely available network protocols/systems are mainly throughput-oriented designs, meeting stringent delay requirements of new applications such as virtual reality and vehicle-to-vehicle communications on cellular network requires new network protocol/system designs. C2TCP is an effort toward that new design direction. 

C2TCP is inspired by in-network active queue management (AQM) designs such as RED and CoDel and motivated by lack of a flexible end-to-end (e2e) approach which can adapt itself to different applications' QoS requirements without modifying any network devices. It copes with unique challenges in cellular networks for achieving ultra-low latency (including highly variable channels, deep per-user buffers, self-inflicted queuing delays, radio uplink/downlink scheduling delays) and intends to satisfy stringent delay requirements of different applications while maximizing the throughput. C2TCP works on top of classic throughput-oriented TCP and accommodates various target delays without requiring any channel prediction, network state profiling, or complicated rate adjustment mechanisms.

We have evaluated C2TCP in both real-world environment and extensive trace-based emulations and compared its performance with different TCP variants and state-of-the-art schemes including PCC-Vivace, Google's BBR, Verus, Sprout, TCP Westwood, and Cubic. Results show that C2TCP outperforms all these schemes and achieves lower average delay, jitter, and 95th percentile delay for packets.
\end{abstract}


%
\IEEEpeerreviewmaketitle

\section{Introduction}
%
%
%
%
\IEEEPARstart{E}{merging} applications such as virtual reality, augmented reality, real time remote health monitoring, automated vehicles and vehicle-to-vehicle communications, real time online gaming, etc. have brought new requirements in terms of latency, throughput, and reliability. These new requirements show a clear need for new designs for the network and its protocols. On the other hand, exponential growth in the cellular networks' traffic during recent years (more than 1200\% over recent five-year period~\cite{mobile-stat}) , due to the advances in cellular network technologies, illustrates the important role of cellular networks in the future Internet. That is why 5G, the next generation of mobile communication technology, holds promise of improved latency, throughput, and reliability.

While nearly all of the today's platfroms are using distributed TCP protocols, J. Jaffe in \cite{jaf} has proved that no distributed congetion control can converge to the operation point in which both the minimum delay and maximum throughput are achieved. This result declares the clear trade-off among throughput and delay for TCP flows. Because e2e delay has been treated less important than throughputs for traditional applications, majority of current TCP protocols are throughput-oriented designs. A simple example is the dominance of Cubic, a loss-based throughput-oriented TCP design, as the default TCP scheme in most of the today's smartphone and PC platforms. However, these throughput-oriented designs cannot support next generation applications with a wide range of delay requirements (e.g. AR/VR applications require less than 20ms delay~\cite{ar_vr}, delay requirement of vehicle-to-vehicle communications can be 5-10ms~\cite{5g_vehicle}, and video streaming/conferencing applications can tolerate delays of 50-100ms). 

Moreover, cellular networks experience highly variable channels, fast fluctuating capacities, self-inflicted queuing delays, stochastic packet losses, and radio uplink/downlink scheduling delays.  These unique characteristics make the problem of achieving low latency and high throughput in cellular networks much more challenging than in wired networks. That is why TCP and its variants (which are mainly designed for wired scenarios) are known to perform poorly in cellular networks~\cite{sprout, ex-tcp, verus, lte_depth, bufferbloat2}.

Inspired by in-network AQM designs such as RED~\cite{red} and CoDel~\cite{codel} and motivated by lack of a flexible e2e approach which can adapt itself to different applications' QoS requirements (without modifying any network devices) and unique challenges in the cellular networks for achieving low latency, we propose C2TCP (\textit{\textbf{C}ellular \textbf{C}ontrolled delay \textbf{TCP}})\footnote{An earlier version of this work titled ``Cellular Controlled Delay TCP (C2TCP)'' was published in IFIP Networking conference~\cite{c2tcp}}. One of the key ideas behind C2TCP's design is to absorb dynamics of unpredictable cellular channels by investigating minimum packet delay in a moving time window. C2TCP works on top of a loss-based TCP such as Cubic~\cite{cubic} and accommodates various target delays. In particular, our contributions in this paper are:
\begin{itemize}
\item Combining the key ideas of in-network AQM schemes and throughput-oriented transport control protocols at end-host to provide a flexible e2e solution which allows applications to choose their level of delay sensitiveness without modifying any network devices (C2TCP only needs to be run on the server side). We showed that achieving good performance does not necessarily comes from complex rate calculation algorithms or complicated channel modelings in cellular networks.\footnote{It is already a known fact that predicting cellular channels is hard ~\cite{verus,3g-channel}}

\item Collecting over 2 hours of cellular traces representing various scenarios and environments in New York City and making them available to the community (detailed in section~\ref{sec_traces}). 

\item Implementing C2TCP in latest Linux Kernel, on top of Cubic, and conducting extensive experiments using both real-world tests and trace-driven evaluations (in a reproducible environment using real-world cellular traces) detailed in sections~\ref{sec_eval_real} and ~\ref{eval}. We have compared performance of C2TCP with several TCP variants (including Cubic~\cite{cubic}, TCP Westwood\cite{west}) and state-of-the-art schemes including PCC-Vivace~\cite{vivace}, Google's BBR~\cite{bbr}, Sprout~\cite{sprout}, and Verus~\cite{verus}. Highlights include: on average, Sprout, Verus, BBR, and PCC-Vivace have $1.94\times$, $3.82\times$, $2.44\times$, and $10.05\times$ higher average delays and $1.64\times$, $3.22\times$, $2.31\times$, and $10.52\times$ higher 95th percentile delays compared to C2TCP, respectively. This great delay performance comes at little cost in throughput. For instance, compared to BBR (which achieves the highest throughput among those 3 state-of-the-art schemes), C2TCP's throughput is only about $20\%$ less.
\end{itemize}

Rest of this paper has been organized as follow: We discuss our main motivations and design decisions in the next section. Section~\ref{sec_design} explains the design of C2TCP and its components. In sections~\ref{sec_anal} and~\ref{sec_why}, we have analyzed C2TCP's steady state behavior and discussed the main reasons behind performance improvements achieved by C2TCP. Sections~\ref{sec_eval_real} and~\ref{eval} include our macro-evaluations in which we have focused on the macro-level performance metrics (delay and throughput). In particular, we present the results of our in-field evaluations in section~\ref{sec_eval_real} and our extensive trace-driven evaluations in section~\ref{eval}.  Later, in section~\ref{sec_eval_micro}, we perform micro-level evaluations and dig deep into the C2TCP's characteristics including TCP friendliness, impact of buffer size on its performance, and compare it with CoDel~\cite{codel}, an AQM scheme that requires modification on carriers network, and show that C2TCP can work very close to CoDel and even in some cases outperform its delay performance.

\section{Motivations and Design Decisions}
\textbf{Flexible e2e Approach:} One of the key distinguishing features of cellular networks is that cellular carriers generally provision deep per-user queues in both uplink and downlink directions at the base station (BS) to increase network reliability~\cite{sprout}. This leads to issues such as self-inflicted queuing delay~\cite{sprout} and bufferbloat~\cite{bufferbloat,bufferbloat2}. A traditional solution for these issues is using AQM schemes like RED~\cite{red}; however, correct parameter tuning of these algorithms to meet the requirements of different applications is challenging and difficult. Although newer AQM algorithms such as CoDel~\cite{codel} can solve the tuning issue, it comes with a new design for the underlining switches and a need of deploying them in the network causes huge CAPEX cost. In addition, in-network schemes lack flexibility. They are based on ``one-setting-for-all-applications'' concept and do not consider that different types of applications might have different delay and throughput requirements. Moreover, with emerging architectures, such as mobile content delivery network (MCDN) and mobile edge computing (MEC)~\cite{mec}, content is being pushed close to the end-users. So, from the latency point of view, the problem of potential large control feedback delay of e2e solutions diminishes if not disappears. Motivated by these shortcomings and new trends, we seek a ``\textit{flexible e2e}'' solution which will let various server applications running inside different systems/virtual-machines/containers (in the cloud/mobile edge) have different desired target delays.

\textbf{Simplicity:} Cellular channels often experience fast fluctuations and widely variable capacity changes over both short and long timescales~\cite{verus}. This property along with several complex lower layer state machine transitions\cite{lte_depth}, complicated interactions between user equipment (UE) and BS \cite{lte-book}, and scheduling algorithms used in BS to allocate resources for users through time which are generally unknown for end-users make cellular channels hard to be predictable if not \textit{unpredictable}~\cite{verus,3g-channel}. These complexities and unpredictability nature of channels motivates us to avoid using any channel modeling/prediction or adding more complexity to cellular networks. We believe that performance doesn't always come from high complexity. Therefore, we seek ``\textit{simple yet effective}'' approaches to tackle the congestion issue in cellular networks.

\textbf{Network as a Black-Box:} In cellular networks, source of delay is vague. The e2e delay could be due to either self-inflicted queuing delay in BS, delays caused by BS' scheduling decisions in both directions, or downlink/uplink channel fluctuations. Although providing feedback from the network to users can guide them to detect the main source of delay and react to it, it needs to have a new design for cellular networks. However, this comes at the CAPEX cost for cellular carriers. Therefore, we will look at the cellular network as a ``\textit{black-box}'' which doesn't directly provide us with any information about itself.

\begin{figure}[!t]
\centering
\includegraphics[width=0.47\textwidth,height=1.7in]{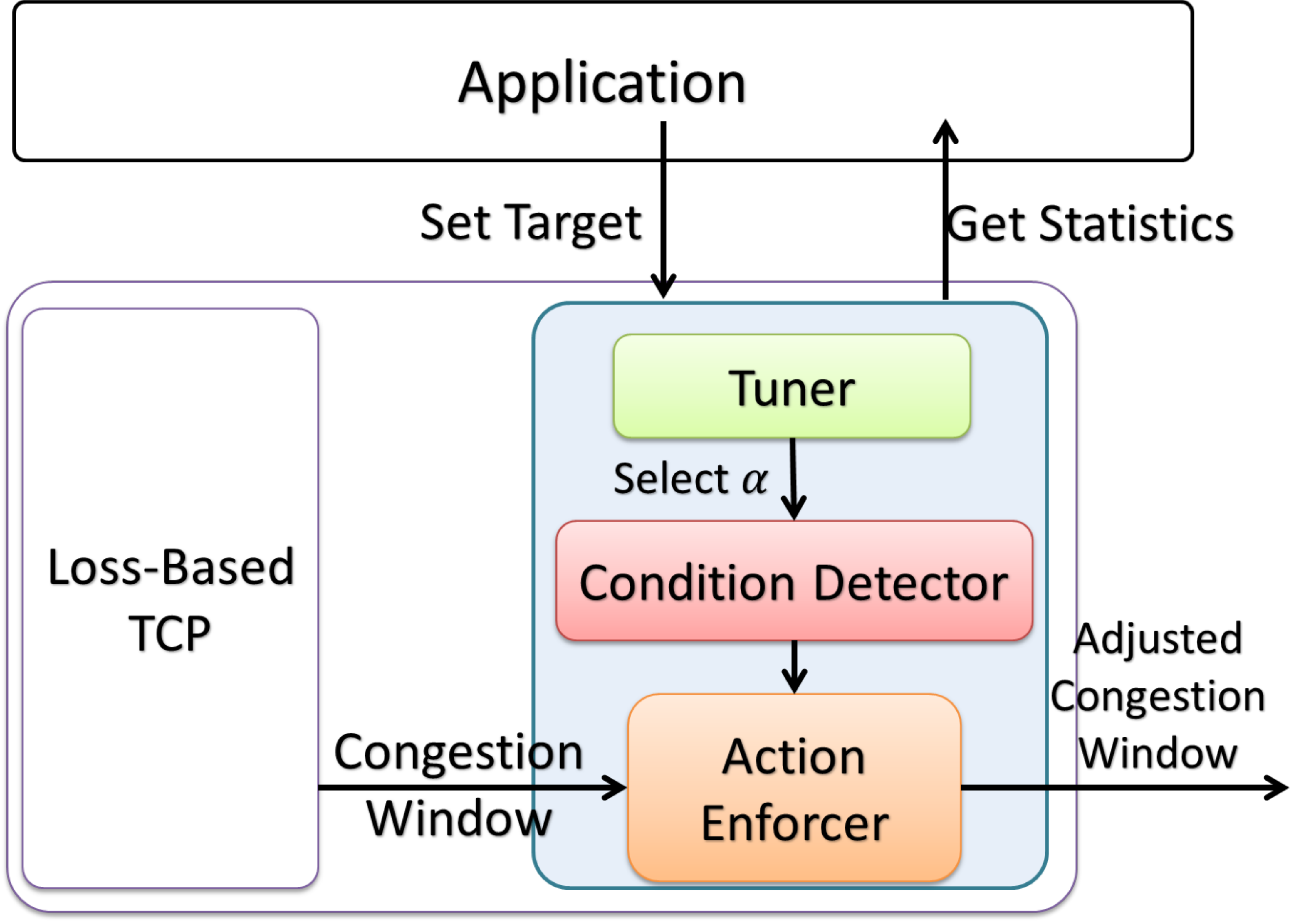}
\caption{A big picture of C2TCP's design}
\label{fig_big-picture}
\end{figure}

\section {C2TCP's Design}
\label{sec_design}
\subsection{Big Picture}
\label{sec_big}
One of the major goals of any TCP scheme is to determine how many inflight packets can exist at different times in the network. This will be done by setting a so-called congestion window (Cwnd) at different times of a TCP session. C2TCP works as an add-on on top of loss-based TCP so that it can inherit the stability, friendliness, and throughput performance of well studied and widely used loss-based approaches such as Cubic. The big picture of the C2TCP's structure on the server side is shown in Fig.~\ref{fig_big-picture}.\footnote{C2TCP only needs to be run on the server and it does not require any changes at the client.} C2TCP consists of two parts: 1- An unmodified loss-based TCP (such as Cubic). 2- A window refining module that runs in parallel with the loss-based TCP in part 1 and refines the Cwnd. The window refining module further consists of three logical blocks: Condition Detector, Action Enforcer, and Tuner.

In a traditional classic loss-based TCP design, it is assumed that an AQM design in the network is responsible to do the queue management and drop the packets if needed\footnote{Let's consider FIFO as a naive AQM approach too}. TCP will then detect the loss and react based on that. Although different AQM designs have different logic, the main idea behind them is the same: They all try to determine when it is a \textit{good condition} in which they can serve packets and when it is a \textit{bad condition} in which they should drop packets. Here, we push this classic observation/assumption further by moving the key responsibility of AQM design (i.e. detecting various network conditions) from the network to the host itself. This design decision will let us have tremendous flexibility compared to a fixed in-network solution. 

More specifically, in C2TCP, Condition Detector acts as an AQM design and tries to figure out the current condition of the network. Based on detected condition, when it is required, Action Enforcer block adjusts the Cwnd of the loss-based TCP. 

The key insight behind the logic of Action Enforcer comes from the following question: \\\textit{If we had an in-network AQM algorithm able to detect a \textit{bad condition}, what would have it done to inform a loss-based TCP and what would be the reaction of the loss-based TCP?}

The answer is a classic straightforward one. It would have simply dropped the packets and caused TCP to time-out (detect the loss) and do a harsh back-off by setting Cwnd to one. So, the key idea of Action Enforcer is to make such an impact by overwriting the Cwnd calculated by loss-based TCP, when condition detector detects a bad condition (an imaginary packet drop).

Meanwhile, the Tuner gets the desired average target delay (called Target) from the application using socket option fields and uses the statistics gathered by C2TCP such as average packet delay to dynamically tune the Condition Detector and increase/decrease its sensitivity for identifying network condition. 

In the following sections the details of each block in Fig.~\ref{fig_big-picture} (Action Enforcer, Condition Detector, and Tuner) and intuition behind their design decisions will be explained.
\begin{figure}[!t]
\centering
\includegraphics[width=0.48\textwidth,height=1.5in]{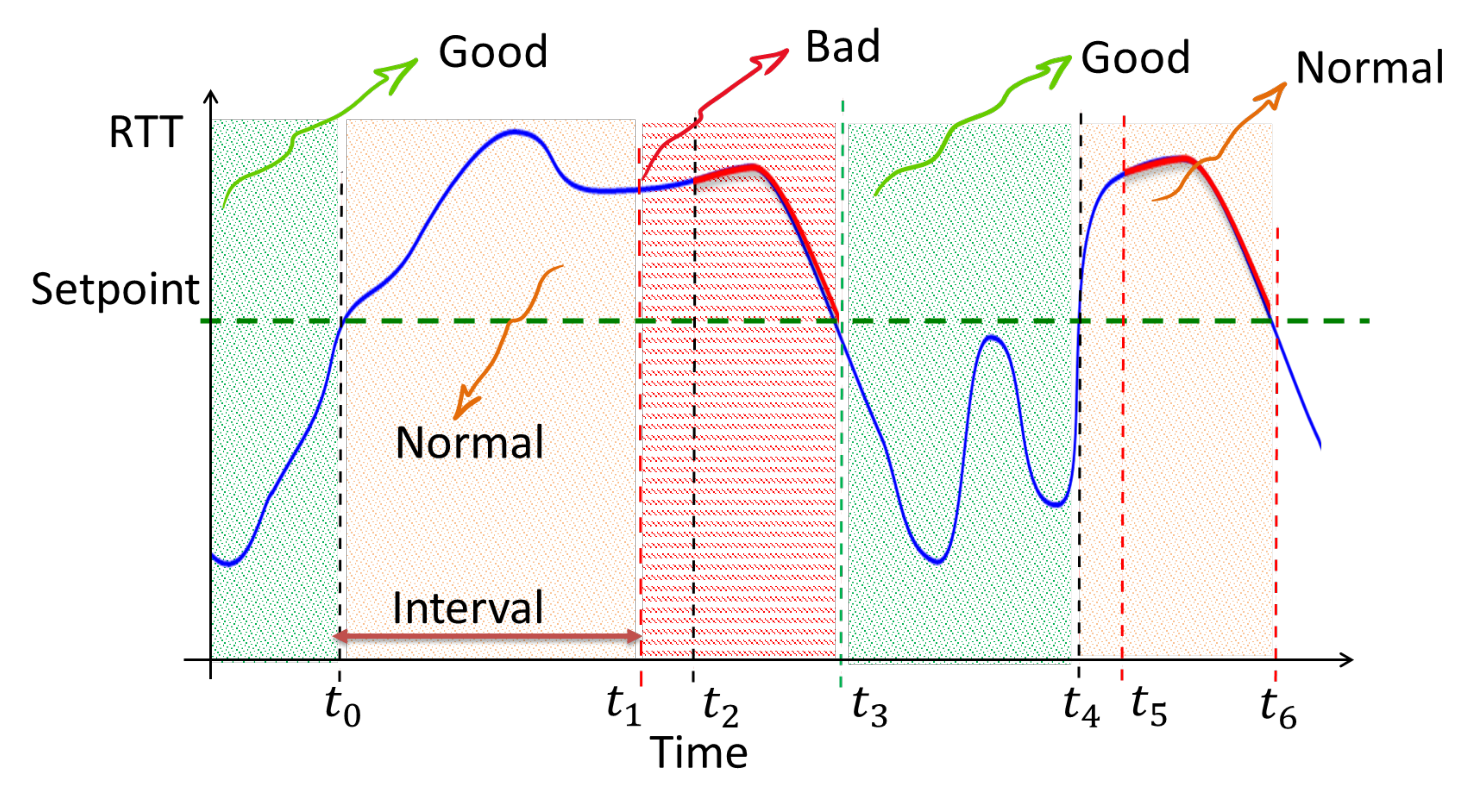}
\caption{The Good, the Normal, and the Bad conditions}\label{fig_good-bad}
\end{figure}
\subsection{Condition Detector}
\subsubsection{The Good, The Normal, and The Bad Conditions}
RTTs\footnote{We use words RTT and e2e delay interchangeably in this paper.} of packets in cellular networks are noisy (due to channel fluctuations, scheduling in uplink/downlink directions at BS, etc.). So, tracing RTT itself won't help to detect network's condition or whether it is congested. However, local minimum RTT observed in a moving time window can work like a filtered version of the noisy RTT. The insight here is that as long as we have a consistent delay for the packets in the network, no matter what the source of delay is (scheduling delay at BS, wireless channel fluctuations, layer 2 packet re-transmission at wireless link, etc.), this dealy can be detected by tracing the minimum RTT in a moving time window (called minRTT from now on). If minRTT is increased across the moving windows, it most likely reflects a consistent delay in the network, while if minRTT is decreased across the moving windows, it most likely shows a good delay response of the network. Using that insight, we define three conditions for the network as follow:
\\We define $\text{RTT}(t)$ as the measured RTT based on an acknowledgment packet received at time \textit{t} and \textit{Interval} as our moving monitoring time window. Now, Given any moment \textit{t} and a desired minRTT called Setpoint, we define: 
\\\textbf{Bad-Condition}: The network is in bad condition, if $min(\text{RTT}(t'))\ge \text{Setpoint}$ for $(t-\text{Interval}) \le t'\le t$.
\\\textbf{Normal-Condition}: The network is in normal condition, if $ \text{RTT}(t) \ge \text{Setpoint}$ and $min(\text{RTT}(t')) < \text{Setpoint}$ for $(t-\text{Interval}) \le t'\le t$.
\\\textbf{Good-Condition}: The network is in good condition, if $\text{RTT}(t) < \text{Setpoint}$.

For instance, consider Fig.~\ref{fig_good-bad} which shows sample RTTs of packets through time. Before $t_0$, RTT of a packet is less than the \textit{Setpoint} value. After $t_0$, RTT goes higher than the \textit{Setpoint} while for any time t, $t_0<t<t_1 (t_1=t_0+\text{Interval})$, $\text{minRTT}(t)<\text{Setpoint}$. So, in $[t_0,t_1]$ the network is in Normal condition. After $t_1$, minRTT goes higher than the \textit{Setpoint} and a Bad-Condition is detected which lasts till $t_3$ when $\text{RTT}(t_3)$ goes below \textit{Setpoint} indicating detection of a Good-Condition. 

Note that the delay responses of packets in $[t_2,t_3]$ and $[t_5,t_6]$ periods are identical. However, since the history of their delay is different at $t_2$ and $t_5$, those two periods have been identified differently (the first one is in a Bad-Condition, while the second one is in a Normal-Condition). This example shows how we can use our definition to qualitatively get a sense of the history of the packet's delay without recording the whole history of delay for all packets. 
\subsubsection{Persistence of Bad Condition}
\begin{figure}[!t]
\centering
\includegraphics[width=0.48\textwidth,height=1.6in]{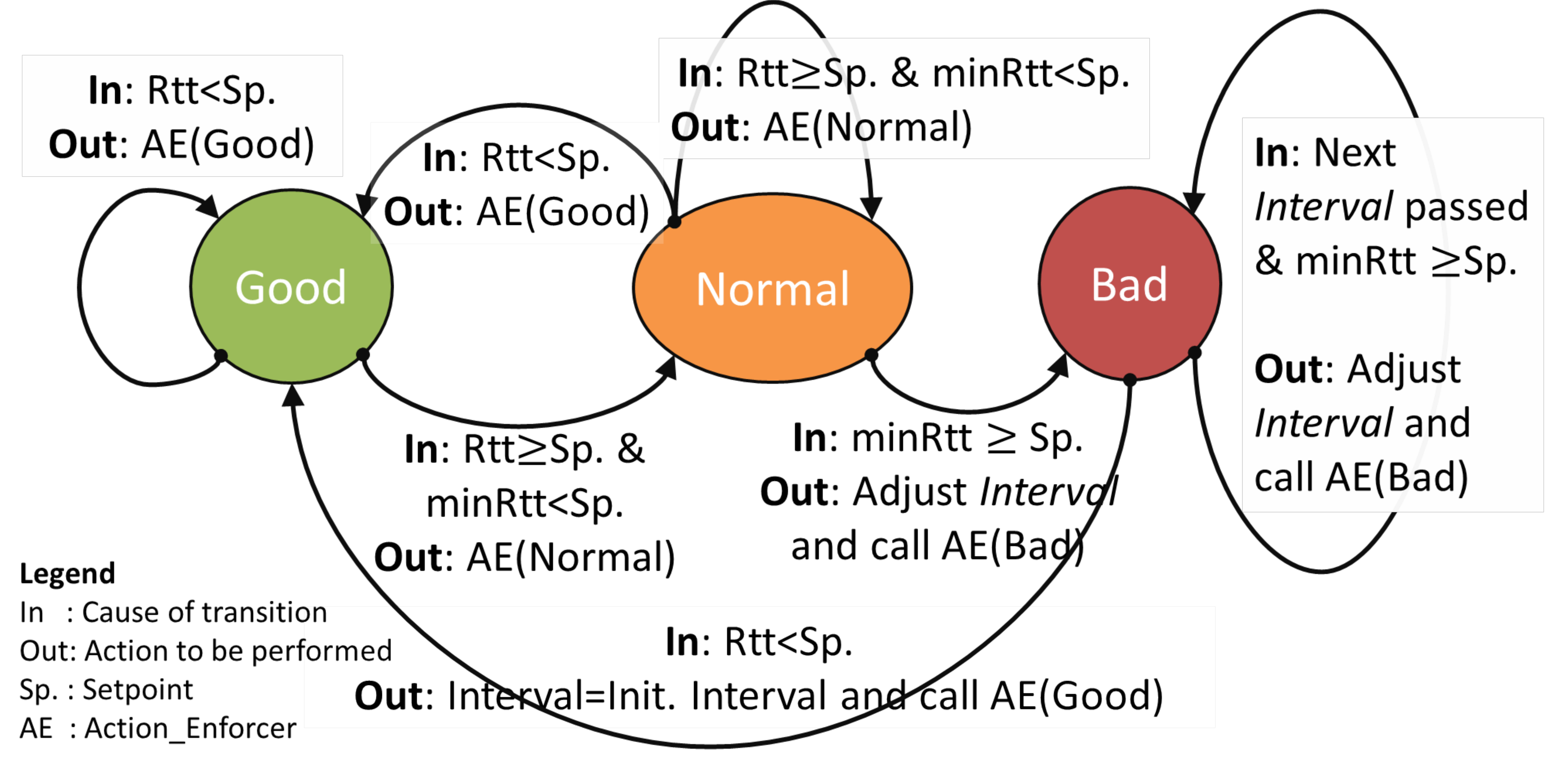}
\caption{State machine of Condition Detector executed upon receiving an Ack packet}\label{fig_fsm}
\end{figure}

Fig.~\ref{fig_fsm} shows the state machine of the Condition Detector and Algorithm~\ref{alg_condition} represents its pseudo code. At the start of each Bad Condition, Action Enforcer will be called to act accordingly. So when a Bad Condition continues, it becomes an alarming situation which requires more frequent reactions from the sender. In other words, when minRTT consistently is high, the sender should proceed with more caution regarding changing its sending rates. Therefore, when a Bad Condition is detected, C2TCP reduces the \textit{Interval} by a factor of $\frac{1}{\sqrt{N}}$ (where \textit{N} is the number of consecutive detected Bad Conditions) and uses the new \textit{Interval} to detect the probable next consecutive Bad-Condition. 

\textbf{Why $\frac{\textbf{1}}{\sqrt{\textbf{N}}}$?} On the one hand, as mentioned in section~\ref{sec_big}, detection of each Bad condition in C2TCP resembles an imaginary packet drop (by an imaginary AQM scheme in the network). In other words, having an increase in the rate of detecting Bad conditions in C2TCP is similar to having an increase in the rate of packet drops by an AQM algorithm. On the other hand, the well-known relationship of drop rate in the network and the loss-based TCP's throughput is that ``the drop rate in the network is proportional to the inverse square of the throughput of loss-based TCP'' (for an example derivation of this relation, check ~\cite{sqrt,sqrt2}). Therefore, similar to ~\cite{codel}, to get a linear decrease in TCP's throughput, we increase the rate of detecting consecutive Bad conditions, which is equal to increasing rate of imaginary packet drops, proportional to the square root of \textit{N}. That explains why we reduce Interval by a factor of $\frac{1}{\sqrt{N}}$.

\textbf{Notice}: As Fig.~\ref{fig_fsm} and Alogithm~\ref{alg_condition} show, during Good Condition, Action Enforcer will be called after receiving every Ack packet. However, Action Enforcer will be called just \textit{at the start} of each Bad Condition. Also, note that as Algorithm~\ref{alg_condition} declares, Condition Detector's implementation intentionally does not require any timer and does not directly calculate local minRTT during each Interval. This greatly simplifies the implementation of Condition Detector.
\begin{figure}[!t]
\small
 \removelatexerror
  \begin{algorithm}[H]
  \DontPrintSemicolon
   \caption{Condition Detector's Logic}
   \label{alg_condition}
    \Fn(\tcp*[h]{process a new received Ack}){pkts\_acked()}{ 
        $rtt \longleftarrow current\_rtt$\;
        $now \longleftarrow current\_time$\;
        $action\_required \longleftarrow false$\;
        \uIf(){$rtt < MINRTT$}{ 
            $MINRTT = rtt$\;
        }
        $Setpoint = \alpha \times MINRTT$\;
        \uIf(){$rtt < Setpoint$}{ 
            $Interval = Setpoint$\;
            $condition = Good$\;
            $first\_time = true$\;
            $N = 1$ \tcp{N: Num. of consecutive backoffs}
            $action\_required = true$\;
        }
        \uElseIf{first\_time}{
            $condition=Normal$\;
            $next\_time = now+Interval$\;
            $first\_time = false$\;
        }    
        \uElseIf(){$now > next\_time$}{
            $condition=Bad$\;
            $next\_time = now+\frac{Interval}{\sqrt{N}}$\;
            $N \mathrel{+}+$\; 
            $action\_required = true$\;
        }
        \uIf(){$action\_required $}{
            Action\_Enforcer($condition,rtt,Setpoint$)
        }
     }
  \end{algorithm}
\end{figure}
\subsubsection{Setpoint and initial Interval values}
In the mechanisms described for detecting network conditions, \textit{Setpoint} and initial \textit{Interval} values are coupled together. When we want to detect minRTT with a Setpoint value of 100ms, if we set initial \textit{Interval} to values way smaller than 100ms, C2TCP will react very fast and report a lot of false Bad-Conditions. In contrast, when we set initial \textit{Interval} to values way larger than 100ms, C2TCP will have a lot of false Good-Conditions. Considering this and the fact that initial \textit{Interval} and \textit{Setpoint} should not be smaller that intrinsic minimum RTT of the network, we choose $initial Interval=Setpoint=\alpha\times MINRTT$ in which MINRTT represents the global minimum RTT (from the start of the session to current time).

The key parameter $\alpha$ determines the level of sensitivity for distinguishing Good and Bad conditions in Condition Detector, the heart of C2TCP. Value of $\alpha$ will be controlled dynamically by Tuner (detailed in section~\ref{sec_tuner}).
\subsection{Action Enforcer}
\label{sec_interval-reduction}
\begin{figure}[!t]
\small
 \removelatexerror
  \begin{algorithm}[H]
  \DontPrintSemicolon
   \caption{Action Enforcer's Algorithm}
   \label{alg_enforcer}
    \Fn(\tcp*[h]{}){Action\_Enforcer($condition$,$rtt$,$Setpoint$)}{ 
        \Switch(){condition}{ 
            \Case{Good}{
                $Cwnd \mathrel{+}= \frac{Setpoint}{rtt}\times \frac{1}{Cwnd}$\;
            }
            \Case(){Normal}{\tcc*[l]{Do nothing!}
            }
            \Case(){Bad}{
                \tcc*[l]{setting ssthresh using default TCP function which normally recalculates it in congestion avoidance phase}
                $ssthresh \longleftarrow recalc\_ssthresh()$\;
                $Cwnd \longleftarrow 1$\;
            }
        }
     }
  \end{algorithm}
\end{figure}
Based on detected condition by Condition Detector, Action Enforcer block adjusts the Cwnd of the loss-based TCP. Action Enforcer's pseudo code is shown in Algorithm~\ref{alg_enforcer}. As discussed in section~\ref {sec_big}, detection of Bad condition is similar to an imaginary drop of packet in the network. So, when Bad Condition is detected at source, Action Enforcer overwrites the decision of loss-based TCP and sets Cwnd to one (similar to having a timeout in loss-based TCP).

In congestion avoidance phase~\cite{cong_avoidance}, loss-based TCP only increases Cwnd by $\frac{1}{Cwnd}$ after receiving each Ack packet so that after one RTT, Cwnd can increase by 1 packet. However, detection of a Good Condition illustrates that there is an opportunity to send more packets into the network and use the current available capacity at a little cost of an increase in self-inflicted queuing delay. So, in Good Condition, in addition to the increase done by the loss-based TCP, Action Enforcer increases the Cwnd so that after one RTT, Cwnd increases by $\frac{Setpoint}{RTT_{current}}$ more packets (equation~\ref{eq_cwnd} (line 4 in Algorithm~\ref{alg_enforcer})). The choice of this additive increase is to follow the well-known AIMD (Additive Increase Multiplicative Decrease) property to ensures that C2TCP's algorithm still achieves fairness among connections~\cite{aimd}. We have examined C2TCP's fairness in more detail in section~\ref{sec_fair}.
\begin{equation} 
\label{eq_cwnd} 
\small
Cwnd_{new} = Cwnd_{recent}+\frac{Setpoint}{RTT_{current}}\times \frac{1}{Cwnd_{recent}}
\end{equation}
When Normal Condition is detected, Action Enforcer performs no further action and doesn't change the Cwnd calculated by the loss-based TCP. Therefore, the loss-based TCP decides the final Cwnd in Normal Condition and adjusts the final Cwnd based on its logic. 
\subsection{Tuner}
\label{sec_tuner}
Delay constraints for various classes of delay-sensitive applications differ from each other. However, the overall picture is that for each class of applications there is usually a desired delay performance. One of the important design features of C2TCP is that it can adapt itself to the application's delay constraint. The Tuner block is responsible for handling this feature. Applications can provide C2TCP with their desired average delay called Target. The Target can be set through a socket option field at the time of creating the TCP socket and can even be changed during the lifetime of that socket.

\begin{figure}[!t]
\centering
\includegraphics[width=0.7\linewidth,height=1.3in]{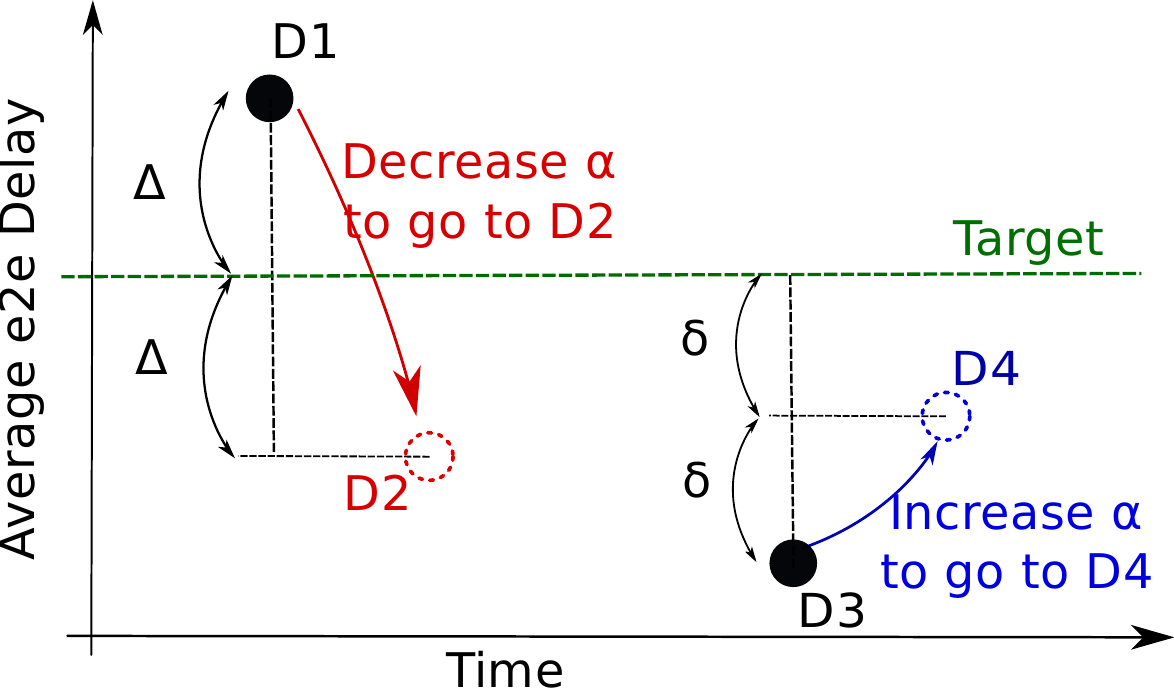}
\caption{Tuner's main logic. }
\label{fig_tuner}
\end{figure}

The Tuner periodically uses the statistics of the average delay of packets and employs the Target given by the application to adjust the $\alpha$ parameter of the Condition Detector block. The overall intuition here is that decreasing $\alpha$ (i.e., decreasing Setpoint) will push Condition Detector toward being more delay-sensitive, while increasing $\alpha$ (i.e., increasing Setpoint) will push it toward being more relaxed and gain higher throughput. The big picture of Tuner's logic and its pseudo code are shown in Fig.~\ref{fig_tuner} and Algorithm~\ref{alg_tuner}, respectively. 

Our sensitivity analysis indicates that a tuning cycle of half a second gives reasonable results, so we used that as our tuning cycle. Tuning cycles much larger than 0.5 seconds will miss channel fluctuations and cause C2TCP to react slowly to them. On the other hand, tuning cycles much smaller than 0.5 seconds make C2TCP too aggressive and could overreact to the good or bad network conditions. This is similar to the observations in \cite{verus} for choosing a profile update rate.

Every 0.5 seconds, Tuner looks into the average delay of packets. Considering Fig.~\ref{fig_tuner}, if the average e2e delay (D1) is bigger than Target, Tuner decreases $\alpha$ (lines 10-13 in Algorithm~\ref{alg_tuner}) to push the average e2e delay of next tuning cycle toward point D2. On the other hand, if average delay is less than Target (D3), Tuner increases $\alpha$ (lines 6-9 in Algorithm~\ref{alg_tuner}) to push the average e2e delay of next tuning cycle toward point D4. As illustrated in Algorithm~\ref{alg_tuner}, the amount of change in $\alpha$ is proportional to the distance of average e2e delay from the desired Target. We will discuss the reasons behind these adjustments in more details in section~\ref{sec_anal}.
\begin{figure}[!t]
\small
 \removelatexerror
  \begin{algorithm}[H]
  \DontPrintSemicolon
   \caption{Tuner's Algorithm}
   \label{alg_tuner}
	\Fn(){Tune()}{ 
		\tcc*[l]{Every 0.5 second tune $\alpha$ using following:}
		$avg\_rtt \longleftarrow$ \textit{average rtt during previous Tuning Cycle}\;
		$min\_\alpha \longleftarrow 1$\;
		$max\_\alpha \longleftarrow 10$\;
		\uIf(){$avg\_rtt<Target$}{ 
			$\alpha\mathrel{+}= \frac{Target-avg\_rtt}{2avg\_rtt}$\;
			\uIf(){$max\_\alpha \leq \alpha$}{ 
				$\alpha=max\_\alpha$\;
			}
		}
		\uElseIf{$Target < avg\_rtt$}{
			$\alpha\mathrel{-}= \frac{2(avg\_rtt-Target)}{Target}$\;
			\uIf(){$\alpha \leq min\_\alpha$}{ 
				$\alpha=min\_\alpha$\;
			}
		}
	 }
  \end{algorithm}
\end{figure}

\section{Analysis of C2TCP's Behavior}
\label{sec_anal}
\begin{figure}[!t]
\centering
	\includegraphics[width=0.85\linewidth,height=1in]{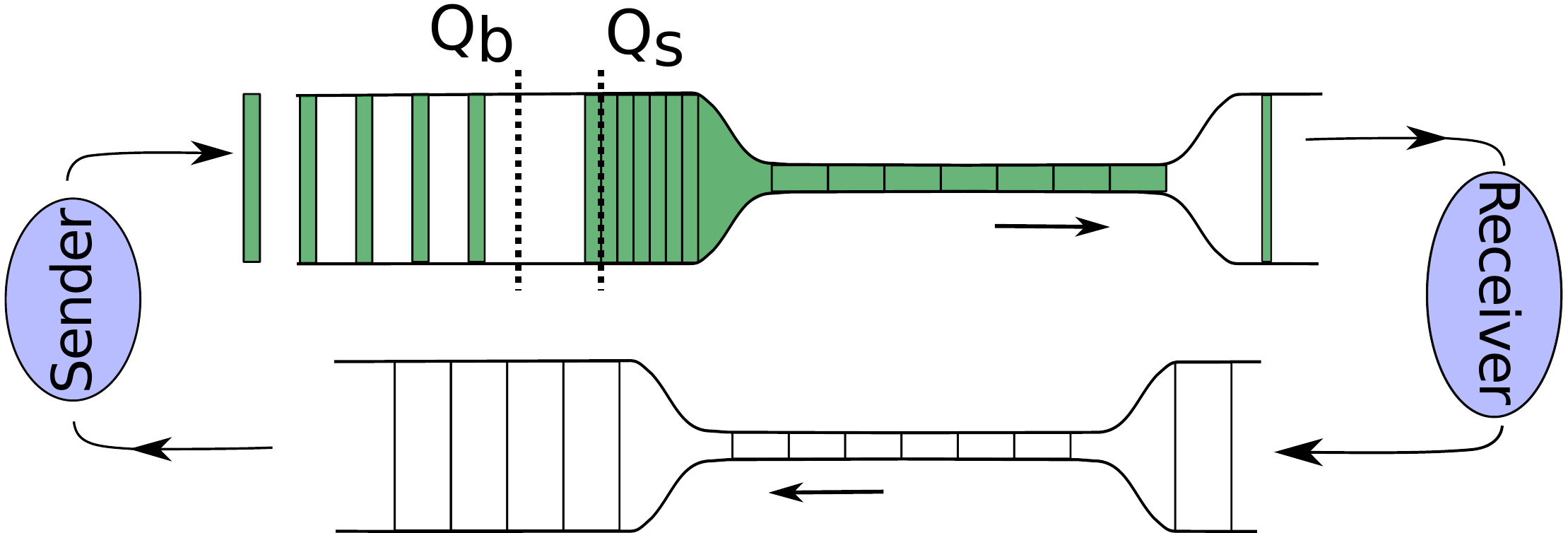}
	\caption{Single queue model of the network}
	\label{fig_q}
\end{figure}

In this section, we use average analysis to investigate the behavior of a single long-lived C2TCP flow and show that C2TCP can upper bound the average e2e delay of packets in the steady state. To simplify the analysis, we model the network as a single queue representing the network's bottleneck link queue as shown in Fig.~\ref{fig_q} where horizontal direction is time, vertical direction shows bandwidth, and green rectangles represent packets. When a packet hits the bottleneck link, it is squeezed in bandwidth and stretched out in time (to have constant area/size). Squeezed out packets reach receiver and receiver sends Ack packets accordingly to the sender. Reception of Acks gives room to the sender for sending more packets. Using that model we define followings:

\textit{$\text{e2eDelay}$}: e2e delay representing the delay between the time of sending a data packet (at sender) and the time of receiving its corresponding Ack packet (at sender)\footnote{Notice: e2e delay consists of propagation delay, transmission delay, and queuing delay in both directions. Propagation delay is usually way smaller than other delays especially when the servers are located at mobile edge network. Moreover, due to the larger size of the data packets compared to the Ack packets (about $25\times$), transmission time of the data packets is way larger than the Ack packets. So, putting all together, e2e delay will be dominated by the delay of the data packets in the downlink direction.}.

\textit{$bw$}: Average pipe\footnote{We use words pipe and bottleneck link interchangeably.} bandwidth (packet per second)

\textit{P}: Number of packets that fill the pipe without causing any queuing delay. Pipe is fully filled without causing queuing delay ($\text{e2eDelay}=\text{MINRTT}$) when arrival rate to the queue (i.e., sender's throughput) is equal to pipe's BW. Therefore:
\begin{IEEEeqnarray}{rCl} 
\label{eq_p} 
&\text{When  }\text{Arrival Rate}=bw\Rightarrow \frac{\text{Cwnd}}{\text{MINRTT}}=bw \Rightarrow \nonumber
\\
&P=\text{Cwnd}=bw\times \text{MINRTT}
\end{IEEEeqnarray}

\textit{$W_s$}: Queuing delay that leads to $\text{e2eDelay}=\text{Setpoint}$ ($W_s=\text{Setpoint}-\text{MINRTT}$)

\textit{$Q_s$}: Corresponding queue length when queuing delay is $W_s$

\textit{$\text{Inflight}$}: Number of inflight packets in the network

\begin{figure}[!t]
\centering
	\includegraphics[width=0.95\linewidth,height=1.7in]{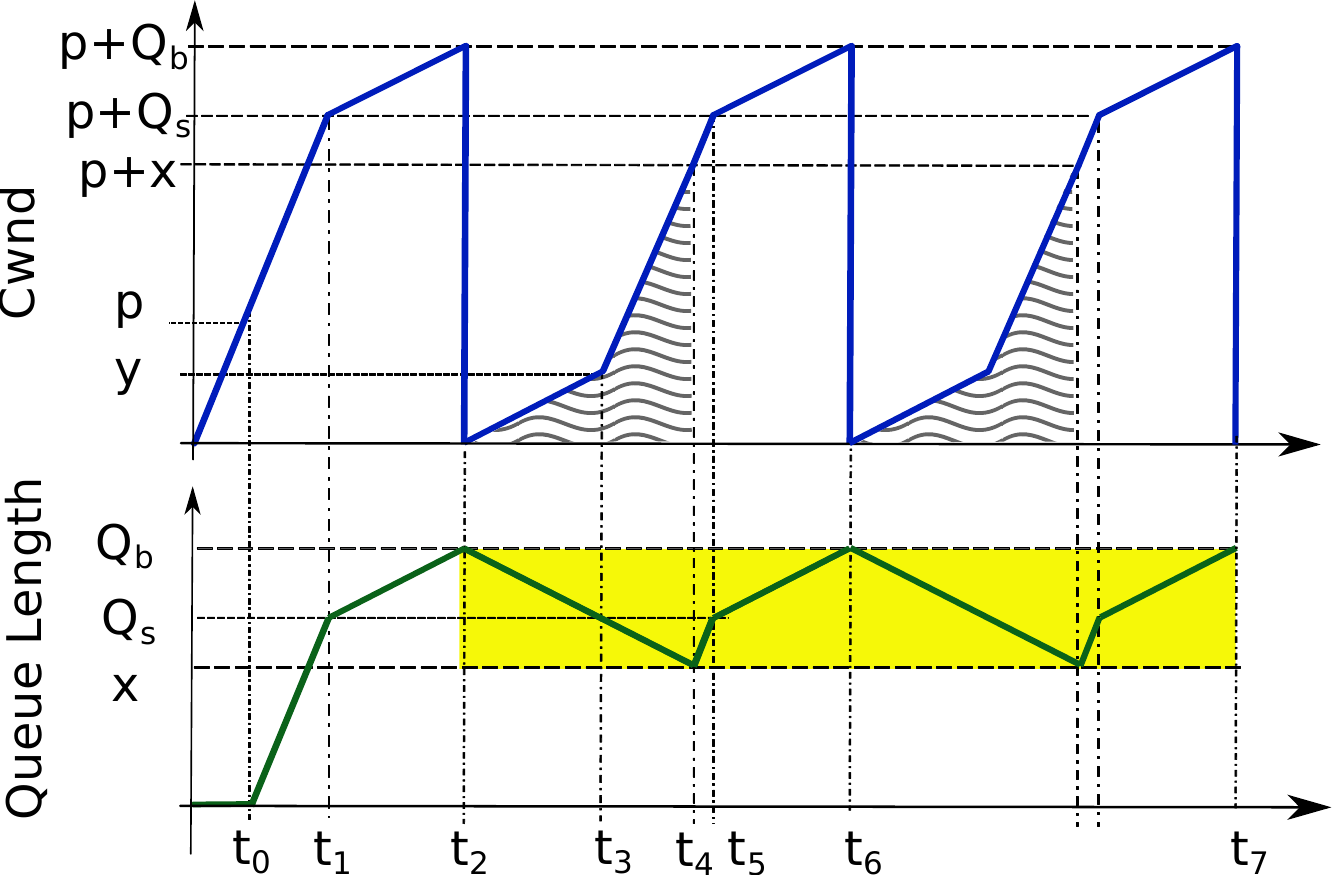}
	\caption{Variations of queue length and Cwnd leading to steady state}
	\label{fig_anal}
\end{figure}
We assume that the startup phase of loss-based TCP has been passed and it is already in congestion avoidance (CA) phase~\cite{cong_avoidance}. Also, to simplify the analysis in steady state, we assume that the average bottleneck link BW ($bw$) remains constant. Now, we follow the variations of Cwnd and queue length shown in Fig.~\ref{fig_anal}\footnote{Without loss of generality, we assume at time 0 (arbitrary time) in Fig.~\ref{fig_anal} queue length is zero.} and calculate the average e2e delay of the C2TCP flow in steady state (we are ignoring the one RTT delay between the time of having a specific queue length and the time of C2TCP’s reaction to that specific queue length since the timescale of the Cwnd growth is long compared to an RTT). Considering Fig.~\ref{fig_anal}: 

\textbf{$[0,t_0)$}: Before $t_0$, since $\text{Cwnd}<P$, there is no queuing delay. Hence, $\text{e2eDelay}=\text{MINRTT}\le \text{Setpoint}$. So, C2TCP is in Good condition and overall Cwnd growth rate is $(1+\frac{\text{Setpoint}}{\text{RTT}}$) packets per RTT (Loss-based TCP increases Cwnd by 1 (in CA phase) and Action Enforcer by $\frac{\text{Setpoint}}{\text{RTT}}$ packets (check equation~\ref{eq_cwnd})).

\textbf{\textit{$[t_0,t_1)$}}: At $t_0$, Cwnd (=Inflight) equals P and queue length starts growing (mismatch of pipe size and Cwnd).

\textbf{\textit{$[t_1,t_2)$}}: At $t_1$, queue length reaches $Q_s$. So, after $t_1$, queuing delay becomes higher than $W_s$ and $\text{e2eDelay}>\text{Setpoint}$. Therefore, at $t_1$, C2TCP detects Normal condition and Cwnd growth rate returns to normal loss-based TCP rate: 1 packet each RTT. Using equation~\ref{eq_p}, $Q_s$ can be calculated as follows:
\begin{IEEEeqnarray}{rCl} 
\label{eq_qs} 
Q_s=W_s\times bw=(\text{Setpoint}-\text{MINRTT})\times bw \nonumber	
\\
=(\alpha-1)\times \text{MINRTT}\times bw=(\alpha-1)\times P
\end{IEEEeqnarray}

\textbf{\textit{$[t_2,t_3)$}}: At $t_2=t_1+\text{Interval}$ (an Interval after $t_1$), C2TCP detects Bad condition, because for the entire period of $(t_1,t_2]$, minimum RTT was larger than Setpoint ($\text{queue Length}>Q_s\Rightarrow \text{queue delay}>W_s \Rightarrow \text{e2eDelay}>\text{Setpoint}$). Therefore, Cwnd is set to 1 by C2TCP. At $t_2$, queue length ($Q_b$) can be calculated as follows:
\begin{IEEEeqnarray}{rCl} 
\label{eq_qb} 
& \left  \{
  \begin{tabular}{c}
  \text{Cwnd growth rate during $(t_1,t_2)=1$ packet per RTT}  \\
  $\text{Setpoint}=\text{Interval}=\alpha\times \text{MINRTT}$
  \end{tabular}
\right . \Rightarrow\nonumber
\\
&  Q_b=Q_s+\text{Interval}/\text{RTT}\le Q_s+\alpha
\end{IEEEeqnarray}
Due to the primary principle of Cwnd-based TCPs, at any arbitrary time, sender only allows to send $(\text{Cwnd}-\text{Inflight})$ packets to the network and when Cwnd is less than Inflight, no new packet will be sent to the network. After detecting Bad condition at $t_2$, we have $\text{Cwnd}=1$ which is smaller than $\text{Inflight}=Q_b+P$. So, no new packet will be sent to the network (wavy patterned areas in Cwnd graph of Fig.~\ref{fig_anal}). Therefore, queue length starts decreasing until it reaches $Q_s$ again at time $t_3$. Using equation~\ref{eq_qb} we have:
\begin{IEEEeqnarray}{rCl} 
\label{eq_tb} 
t_3-t_2=\frac{Q_b-Q_s}{bw}\le \frac{\alpha}{bw}
\end{IEEEeqnarray}
Druing $[t_2,t_3)$, although Action Enforcer does not increase Cwnd, loss-based TCP still increases it at the rate of 1 packet per RTT. So, at $t_3$, using equations~\ref{eq_tb},~\ref{eq_p}, and~\ref{eq_qs} and the fact that ($P>1$)\footnote{P is generally way larger than 1 (e.g., for a typical LTE network P=40ms$\times$25Mbps=250 packet per sec (for packet size=500B).}:
\begin{IEEEeqnarray}{rCl} 
\label{eq_t3-a1} 
& \left  \{
  \begin{tabular}{c}
	$\text{Cwnd}=1+\frac{t_3-t_2}{\text{RTT}}\le 1+\frac{\alpha}{\text{RTT}\times bw}\le 1+\frac{\alpha}{P}$\\
	$\text{Inflight}=Q_s+P=\alpha \times P$
  \end{tabular}
\right . \Rightarrow\nonumber
\\
\label{eq_t3} 
&\text{Cwnd}<\text{Inflight}
\end{IEEEeqnarray}
The important result from equation~\ref{eq_t3} is that no new packet is sent to the network during $(t_2,t_3)$. This means that queue length will come below $Q_s$ after $t_3$. 

\textbf{\textit{$[t_3,t_4)$}}: At $t_3$, queue length becomes $Q_s$ and $\text{e2eDelay}=\text{Setpoint}$. So, C2TCP enters the Good condition and Cwnd growth rate increases (equation~\ref{eq_cwnd}). At $t_4$, where $\text{Cwnd}=\text{Inflight}$, C2TCP starts sending new packets which causes queue length to increase.

\textbf{\textit{$[t_4,t_5)$}}: Increase of queue length continues. 

\textbf{\textit{$[t_5,t_6)$}}: During $[t_5,t_6)$, C2TCP behaves similar to $[t_1,t_2)$. So, at $t_6$, a Bad condition is detected, Cwnd is set to 1, and C2TCP's behavior during $[t_2,t_6]$ repeats.

Now, we drive the average RTT of packets during steady state ($[t_2,t_6]$). Average queue length of $[t_2,t_6]$ is equal to average of the area under queue length curve in that period. By inspection, we have:
\begin{IEEEeqnarray}{rCl} 
\label{eq_avg} 
& Q_{avg}<\frac{\frac{(Q_b-Q_s)(t_2-t_1)}{2}+Q_s(t_6-t_2)+\frac{(Q_b-Q_s)(t_6-t_5)}{2}}{(t_6-t_2)} \nonumber
\\
& \Rightarrow Q_{avg}<\frac{Q_b-Q_s}{2}+Q_s \text{ (using equation~\ref{eq_qb})} \Rightarrow \nonumber
\\
& Q_{avg}<Q_s+\frac{\alpha}{2} 
\end{IEEEeqnarray}
Now, using equations~\ref{eq_avg},~\ref{eq_qs}, and~\ref{eq_p}:
\begin{IEEEeqnarray}{rCl} 
\label{eq_rtt-avg} 
& RTT_{avg}=\text{MINRTT}+\frac{Q_{avg}}{bw}<\text{MINRTT}+\frac{Q_s+\frac{\alpha}{2}}{bw} \nonumber
\\
& \Rightarrow RTT_{avg}<\text{MINRTT}+(\alpha-1)\text{MINRTT}+\frac{\alpha}{2bw}\Rightarrow \nonumber
\\
& RTT_{avg}<\alpha (\text{MINRTT}+\frac{1}{2bw})
\end{IEEEeqnarray}
Equation~\ref{eq_rtt-avg} shows an upper bound for $RTT_{avg}$ using C2TCP (in steady state).
A more relaxed upper bound can be derived by considering $\text{MINRTT}>\frac{1}{bw}$:
\begin{IEEEeqnarray}{rCl} 
\label{eq_rtt-relax} 
RTT_{avg}<1.5\alpha \text{MINRTT}=1.5 \times\text{Setpoint}
\end{IEEEeqnarray}
Although equations~\ref{eq_rtt-avg} and ~\ref{eq_rtt-relax} declare steady state upperbounds of $RTT_{avg}$, channel link fluctuations and scheduling delay variations can still impact the $RTT_{avg}$ (non steady states). That is why Tuner block will always try to adjust the $\alpha$ parameter so that long-term $RTT_{avg}$ remains less than the desired application Target. More specifically, after one tuning cycle with $\text{Setpoint}=Setpoint_1$, if $RTT_{avg}>\text{Target}$, then to cancel out increase in $RTT_{avg}$ and to push average e2e delay toward $RTT'_{avg}=RTT_{avg}-2(RTT_{avg}-\text{Target})$ in the next cycle (Point D2 in Fig.~\ref{fig_tuner}), Tuner decreases Setpoint (smaller upper bound in equation~\ref{eq_rtt-relax}) using equation~\ref{eq_tuner} (line 11 in Algorithm~\ref{alg_tuner}). On the other hand, if $RTT_{avg}<\text{Target}$, after one tuning cycle, Tuner tries to conservatively increase throughput by compensating e2e delay and push e2e delay toward $RTT'_{avg}=\text{Target}-(\frac{\text{Target}-RTT_{avg}}{2})$ in the next cycle (Point D4 in Fig.~\ref{fig_tuner}). So, Tuner increases Setpoint (larger upper bound in equation~\ref{eq_rtt-relax}) using equation~\ref{eq_tuner} (line 7 in Algorithm~\ref{alg_tuner}).

\small
\begin{IEEEeqnarray}{rCl} 
\label{eq_tuner}
&RTT_{avg}<1.5 Setpoint_1 \Rightarrow RTT'_{avg}<1.5(\frac{RTT'_{avg}}{RTT_{avg}}Setpoint_1)\nonumber
\\ 
&\Rightarrow Setpoint_{new}=\frac{RTT'_{avg}}{RTT_{avg}}Setpoint_1
\end{IEEEeqnarray}\normalsize
\section{Why It Works}
\label{sec_why}
\begin{figure}[!t]
\centering
    \includegraphics[width=0.48\textwidth,height=1.5in]{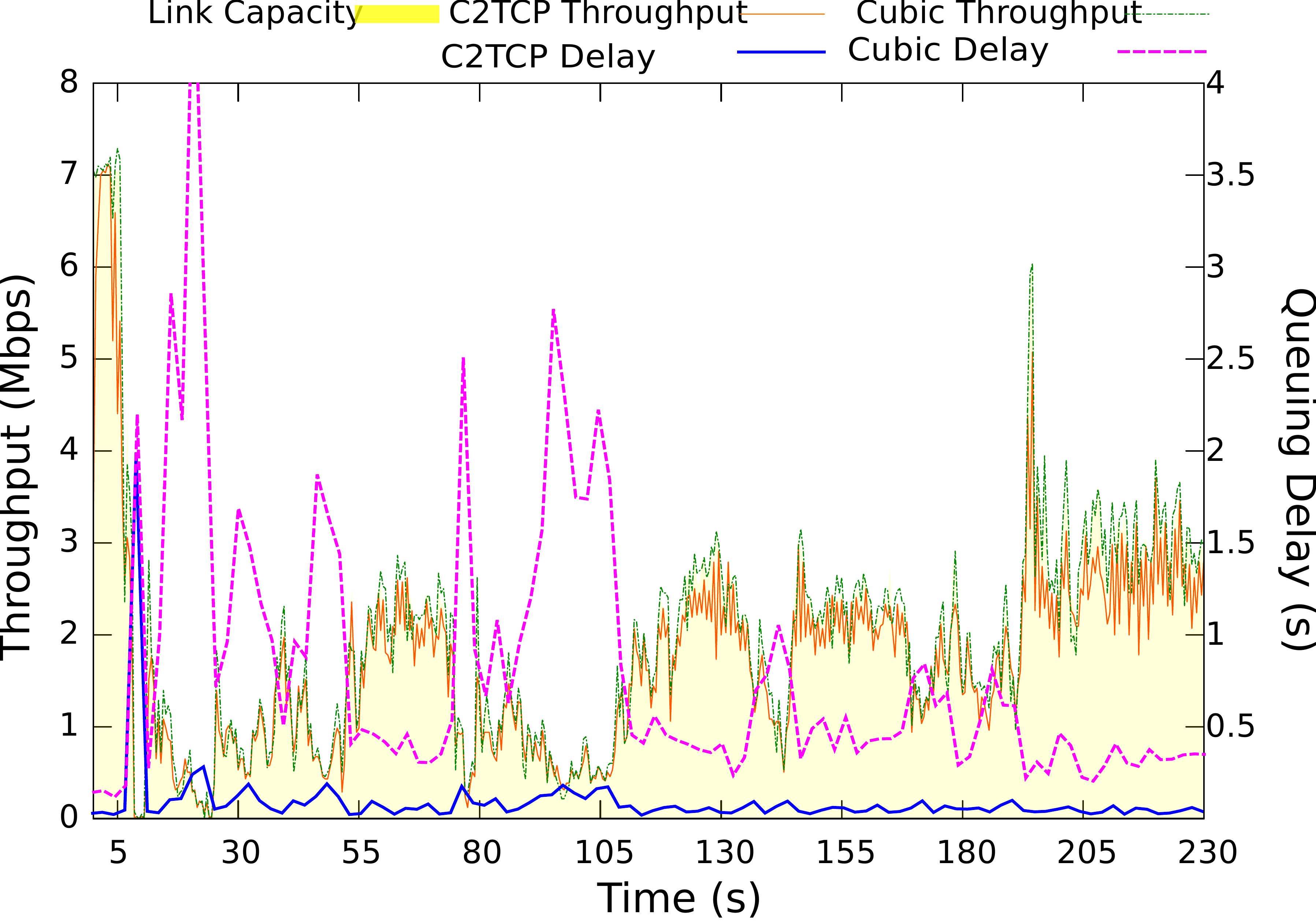}
    \caption{Delay and throughput performnace of Cubic and C2TCP}
    \label{fig_why}
\end{figure}

To show the improvements achieved by C2TCP and highlight the reasons, we compare the performance of C2TCP implemented on top of Cubic with Cubic alone following instructions described in section~\ref{eval}. Here, Target is set to 50ms. Fig.~\ref{fig_why} shows about 2 minutes of varying capacity of a cellular link (TMobile UMTS network in downlink direction measured by prior work~\cite{sprout}) and delay/throughput performance of C2TCP and Cubic.

\textbf{Avoiding excessive packet sending}: Due to variations in link capacity and deep per-user buffers, Cubic's delay performance is poor, especially when there is a sudden drop in link capacity after experiencing good capacity (for instance, look at $[5s-25s]$ and $[70s-80s]$ time periods in Fig.~\ref{fig_why}). However, C2TCP always performs very well regardless of the fast link fluctuations. As results of the analysis in section~\ref{sec_anal} indicate, the key reason is that C2TCP always keeps \textit{proper} amount of packets in the queues so that on the one hand, it avoids queue buildup and increase in the packet delay and on the other hand, it achieves high utilization of the cellular access link when either channel quality becomes good or BS' scheduling algorithm allows serving packets of the corresponding UE.\footnote{When there is no signal or when downlink capacity is close to zero (e.g. $[7s-10s]$ and $[20s-25s]$ in Fig.~\ref{fig_why}), any algorithm including C2TCP will experience delay.} 

\textbf{Absorbing dynamics of channel}: Monitoring minimum RTT in a moving time window allows C2TCP absorb dynamics of cellular link's capacity, scheduling delays, and in general, different sources of delay in network, without a need for having knowledge about the exact sources of those delays, which in practice, are hard to know at end-hosts. 

\textbf{Cellular link as the bottleneck}: Based on high demand of cellular-phone users to access different type of contents, new trends and architectures such as MEC~\cite{mec}, MCDN (e.g.~\cite{maxcdn}), etc. have been proposed and used recently to push the content close to the end-users. So, cellular access link known as the \textit{last-mile} becomes the bottleneck even more than before. This trend helps C2TCP's design to concentrate on the delay performance of the last-mile and boost it.

\textbf{Isolation of per-user queues in cellular networks}: Since C2TCP targets cellular networks, it benefits from their characteristics. One of the important characteristics of cellular networks is that usually different UEs get their own isolated deep queues at BS and there is rare competition for accessing queue of one UE by flows of other UEs~\cite{verus,sprout,bufferbloat2}. This property puts BS' scheduler in charge of fairness among UEs using different algorithms such as weighted round robin, or proportional fairness. This fact helps C2TCP to focus more on the delay performance and leave the problem of maintaining fairness among UEs on the last-mile to the scheduler. In addition, it is usually one critical flow for each UE. C2TCP benefits from this fact too. \footnote{If not, users can simply prioritize their flows locally, and send/request the highest priority one first.}

\textbf{What if C2TCP shares a queue with other flows}: Although the main bandwidth bottleneck in cellular networks is the last-mile, there still might be concern about the congestion before the last-mile access link (for instance, in the carrier's network). The good news is that in contrast with large queues used at BS, normal switches and routers use small queues~\cite{buffer-size}. So, using well-known AIMD property ensures that the C2TCP will achieve fairness across connections~\cite{aimd} before the flow reaches its isolated deep buffer at BS. In section~\ref{sec_fair}, we show good fairness property of C2TCP in the presence of other flows in such a condition.
\begin{figure*}[!t]
\centering
	\begin{minipage}[b]{0.48\linewidth}
			\centering
		\includegraphics[width=0.9\textwidth,height=1.5in]{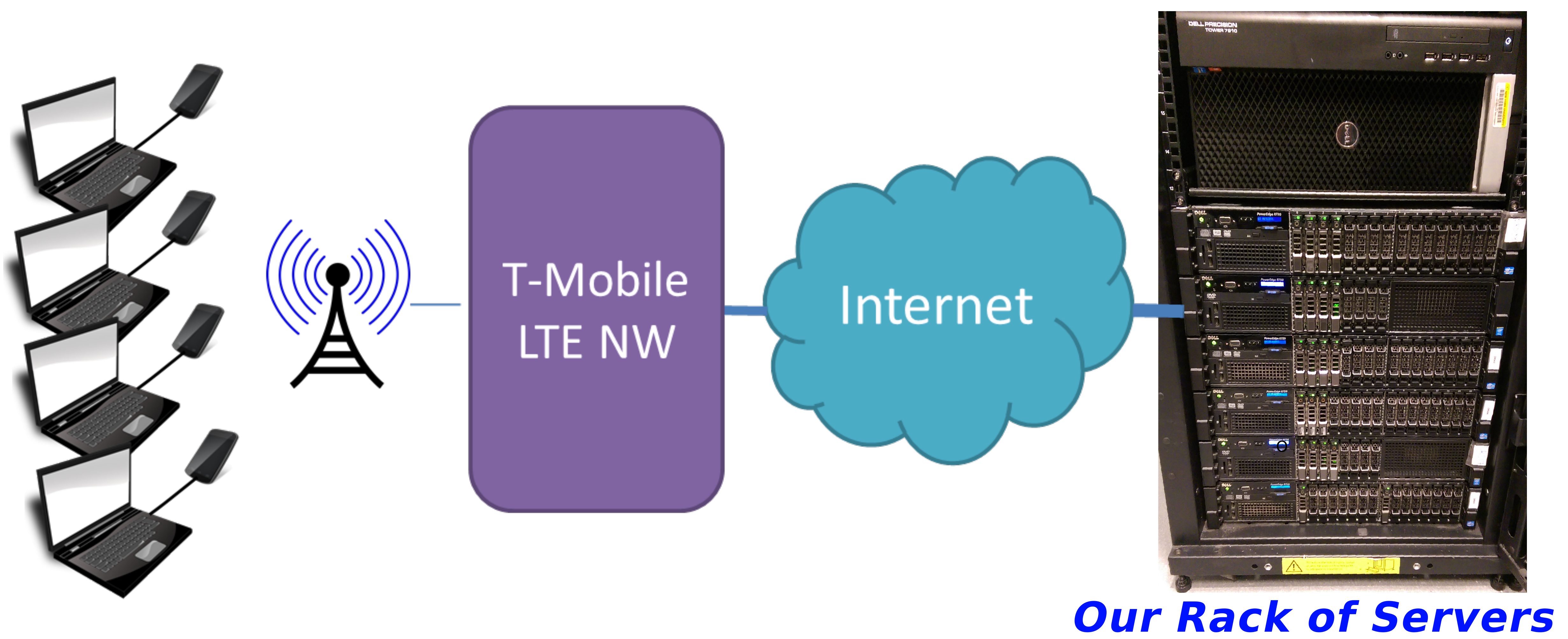}
		\caption{The topology used for real-world evaluations}
		\label{fig_topo_real}
	\end{minipage}		
	\begin{minipage}[b]{0.48\linewidth}
			\centering
		\includegraphics[width=0.9\textwidth,height=1.65in]{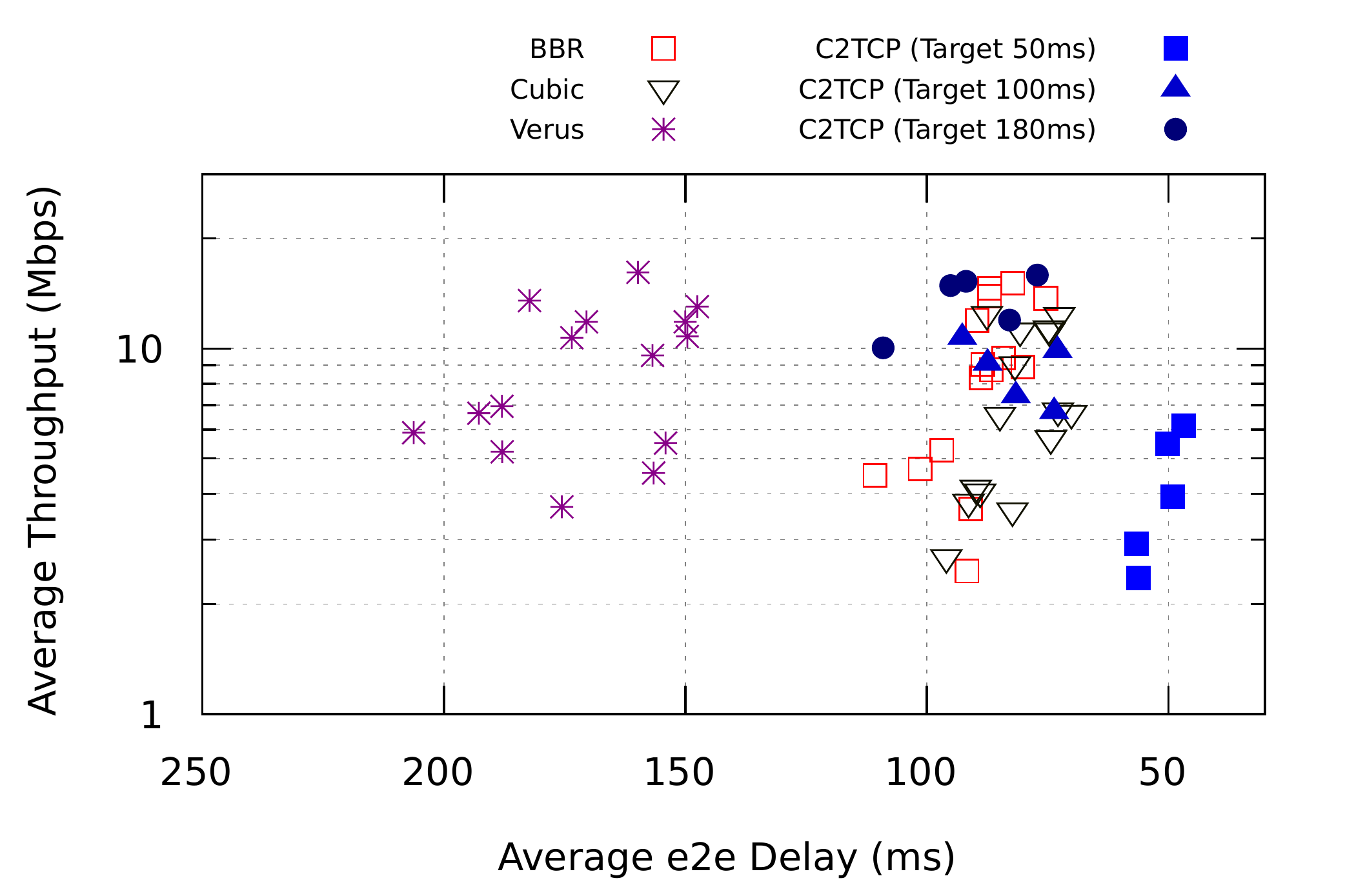}
		\caption{Average throughput and e2e delay for C2TCP, Cubic, Verus, and BBR for all 	experiments}
		\label{fig_results_real_all}
	\end{minipage}
\end{figure*}

\textbf{Letting loss-based TCP do the calculations}: Another helpful insight behind C2TCP is that in contrast with delay-based TCPs, C2TCP does not directly use the delay of packets to calculate the congestion window, but let loss-based TCP, which is basically designed to achieve high throughput~\cite{cubic,reno,newreno,tahoa}, do most of the job. So, instead of reacting directly to every large RTT, the idea of identifying Bad Condition helps C2TCP detect persistent delay problems in a time window and react only to them. 
\begin{figure}[!t]
\centering
	\begin{minipage}[b]{0.48\linewidth}
			\centering
		\includegraphics[width=0.9\textwidth,height=1.3in]{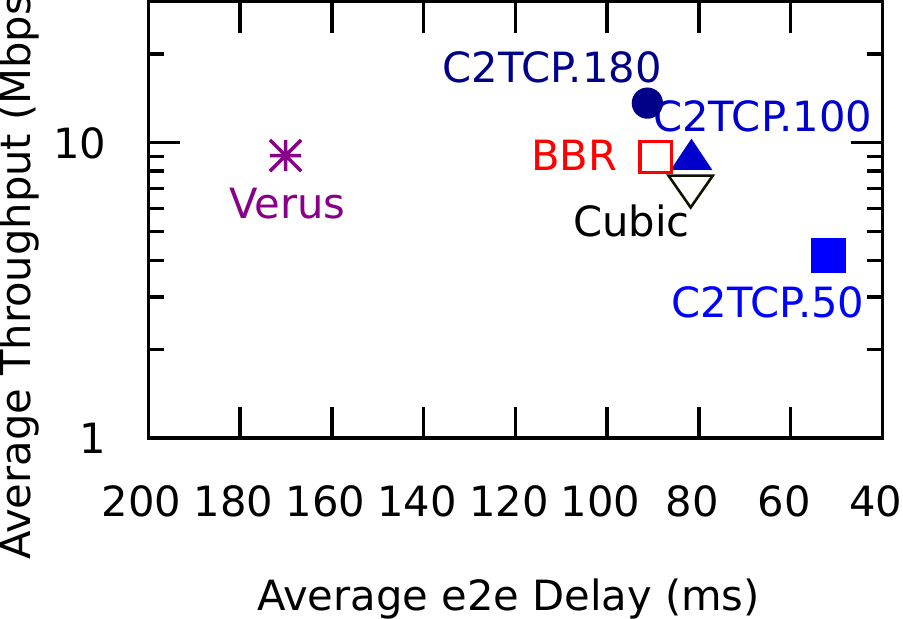}
		\caption{Overall averaged throughput and delay}\label{fig_results_real_avg}
	\end{minipage}		
	\begin{minipage}[b]{0.48\linewidth}
			\centering
		\includegraphics[width=0.9\textwidth,height=1.3in]{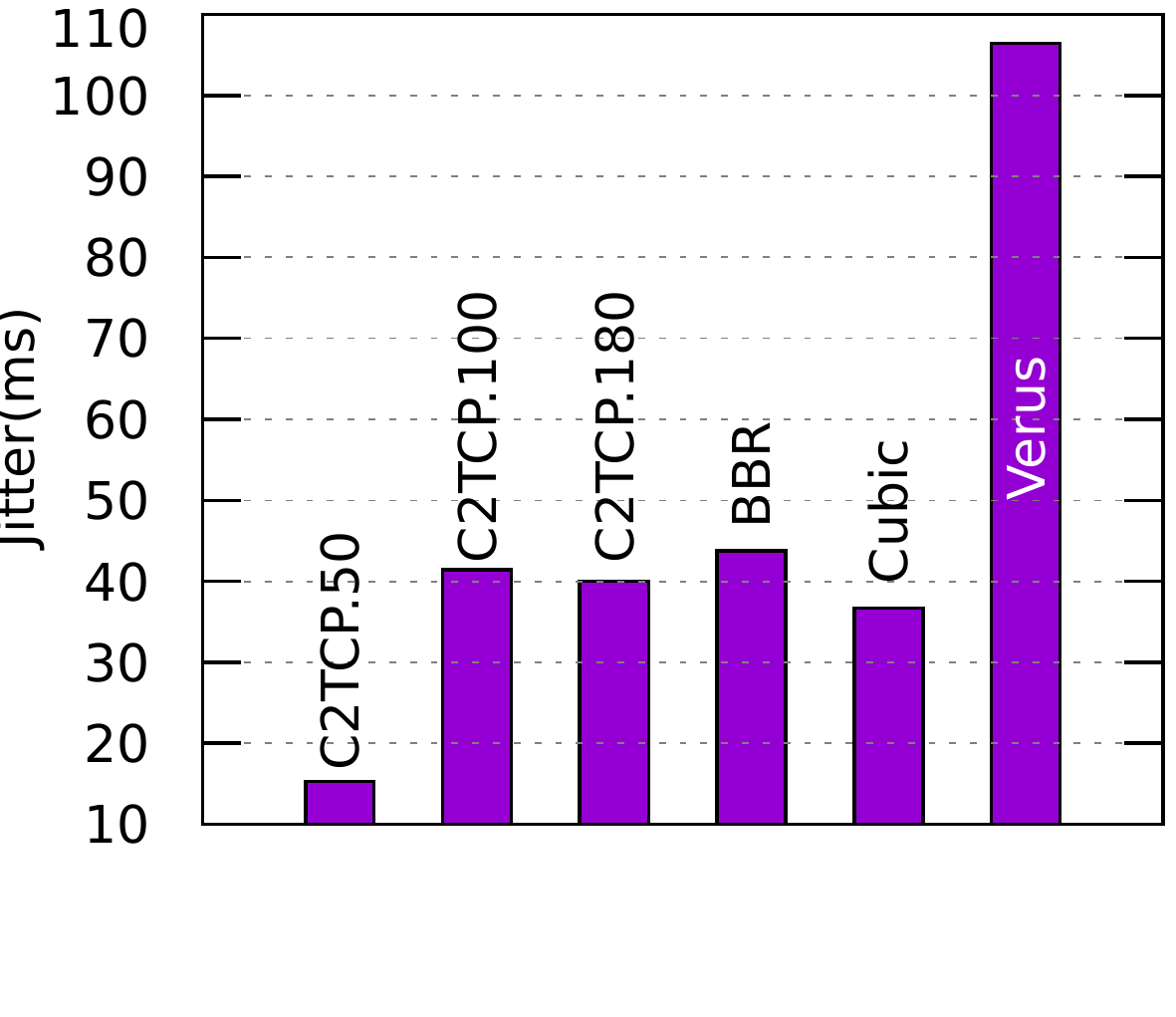}
		\caption{Overall averaged jitter}\label{fig_results_real_jitter}
	\end{minipage}
\end{figure}
\section{In-Field Evaluations}
\label{sec_eval_real}
In this section, we evaluate the performance of C2TCP in the real-world environment by considering three main macro-level performance metrics: delay, jitter, and throughput. 
We have implemented C2TCP in Linux Kernel 4.13, on top of Cubic as the base loss-based TCP and used this implementation in all experiments (source code is available to the community at: \url{https://github.com/soheil-ab/c2tcp}). We perform our real-world evaluations on T-Mobile LTE network in New York City and use 4 Motorola Moto E4 smartphones in our tests. To make sure that all phones are connected to the BS using the same band, we force all of them to use the LTE B4 band. The topology of real-world tests is depicted in Fig.~\ref{fig_topo_real}. We tether phones' LTE connections to laptops and use them as our LTE clients. Moreover, we use 4 servers equipped with very high bandwidth and very low network latency located in the same rack in our lab to send traffic to the clients. Having them in the same rack reduces the chance of having different path latency and throughput for different server-client connections. 

To emulate contention and having competing traffic at BBU, we simultaneously send three UDP streams from three different servers to three clients and at the same time start sending TCP traffic from a server to one of the clients under the test. Specifically, we consider the following scenarios:
\begin{enumerate}
\item Three servers each sending a UDP stream and one server sending C2TCP flow (C2TCP's client-server application)  
\item Three servers each sending a UDP stream and one server sending Google's BBR flow (Iperf3 traffic)
\item Three servers each sending a UDP stream and one server sending Cubic flow (Iperf3 traffic)
\item Three servers each sending a UDP stream and one server sending Verus flow (Verus' client-server application)
\end{enumerate}
We use three different Target values (50, 100, 180 ms) for C2TCP experiments\footnote{minimum RTT in this setup is around 20ms}. For each Target value, we run above scenarios for 30 seconds and repeat each test 5 times. Moreover, we make sure that all phones have the same quality of channel and connected to the same BS during the experiment. All tests are done at the same time and at the fixed location (evening in stationary position in a residential building). Average e2e delay and average throughput for all experiments are shown in Fig.~\ref{fig_results_real_all}. The overall averaged throughput and e2e delay and overall averaged jitter (defined as mean deviation (smoothed absolute value) of delay) over all runs are shown in Fig.~\ref{fig_results_real_avg} and Fig.~\ref{fig_results_real_jitter}, respectively.

As Fig.~\ref{fig_results_real_avg} shows, C2TCP can control the average delay of packets based on the Target very well. As expected, increasing Target decreases the average delay performance, while it allows the sender to achieve higher throughput performance. Also, increasing Target to larger values will push C2TCP toward the performance of Cubic and as expected, make it more throughput hungry. C2TCP's jitter performance follows the same pattern. As Fig.~\ref{fig_results_real_jitter} shows, C2TCP can achieve very low jitters for small Target delays such as 50ms and its jitter performance becomes similar to jitter performance of Cubic when large Targets are selected.
\section{Trace-Driven Macro-Evaluation}
\label{eval}
Here, we evaluate the performance of C2TCP using extensive trace-driven emulation and compare its performance with existing protocols under a reproducible network condition. We use Mahimahi~\cite{mahi} as our trace-driven emulator. 
\subsection{Cellular Traces} 
\label{sec_traces}
\begin{figure}[!t]
\centering
    \begin{minipage}[b]{0.48\linewidth}
        \centering
        \includegraphics[width=\textwidth,height=1.1in]{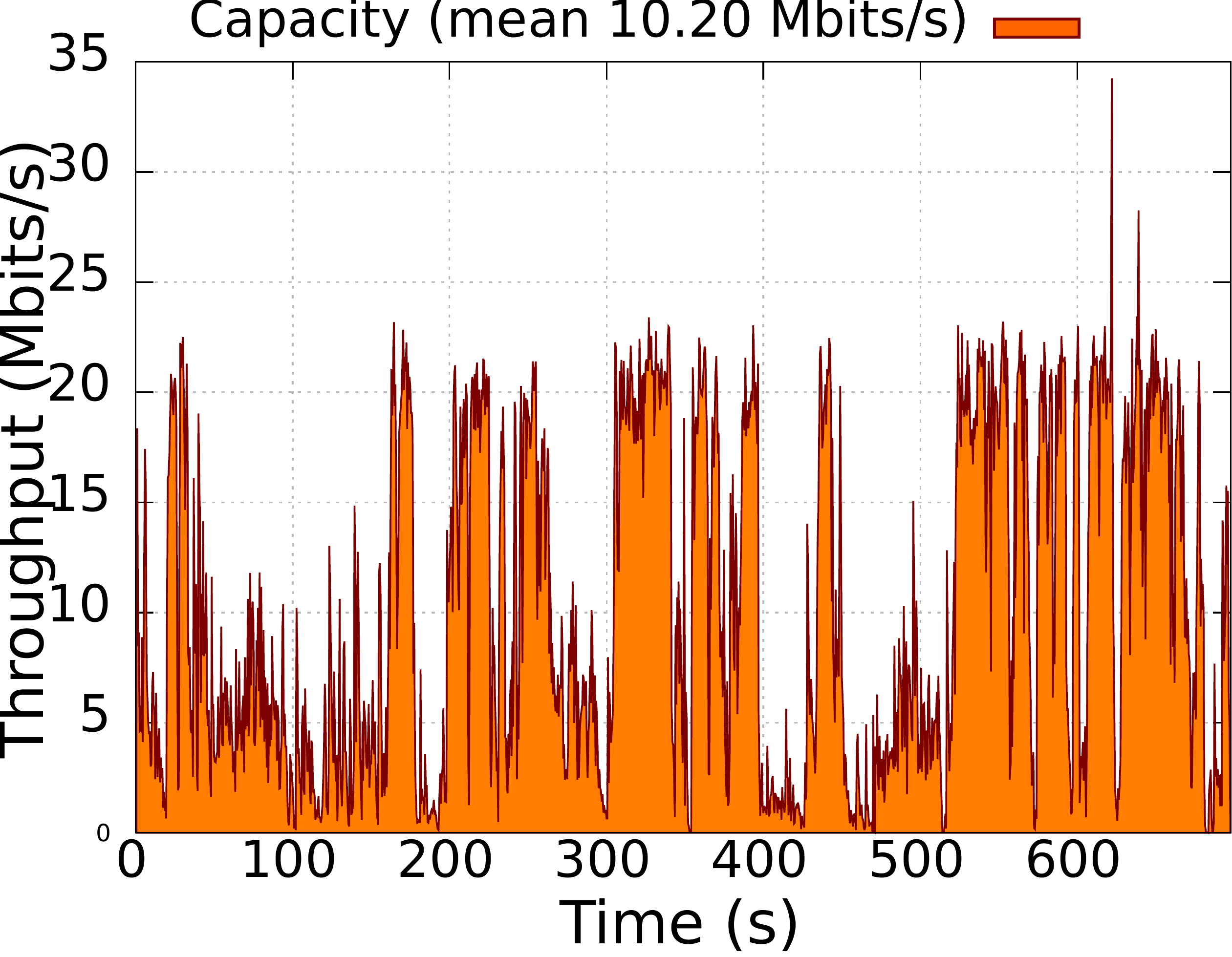}     
        \caption{Cellular downlink traces collected on a subway with cross traffic using LTE technology}        
        \label{fig_trace_subway_cross}
    \end{minipage}
    \hfill\hfill
    \begin{minipage}[b]{0.45\linewidth}
        \centering
        \includegraphics[width=\textwidth,height=1.1in]{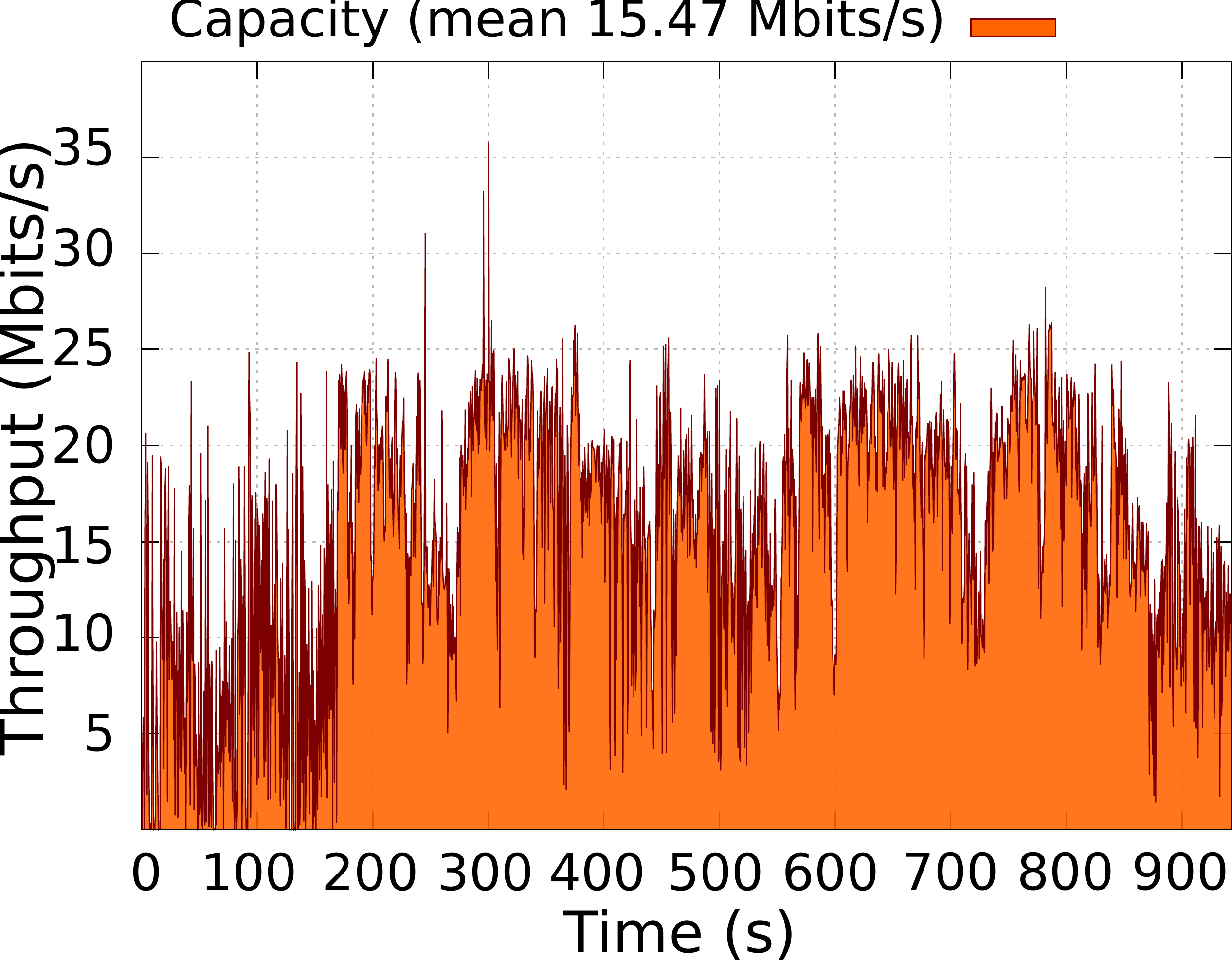}
        \caption{Cellular downlink traces collected at Times Square without cross traffic using LTE technology}
        \label{fig_trace_times}
    \end{minipage}
\end{figure}

To cover the wide range of environments, we have collected 8 new traces using the traffic generator tool (Saturator) provided by prior work\cite{sprout} on T-Mobile network in New York City. We have considered two main scenarios: 1-When UE is moving, and 2-When UE is not moving but the environment is changing. For the moving Scenario, we have collected traces when riding a subway in New York City. For the second scenario, we have collected traces in the stationary position in one of the most crowded places in the world, Times Square. For each scenario, we have considered two cases. In the first case, we send traffic between one of our servers in our lab and one UE in both downlink and uplink directions (minimum RTT between the server and the UE is 20ms). In other cases, to collect the impact of competing traffics at BS on a specific client, we send traffic between another server located in our lab and another client which is located beside the client under the test. We have repeated all measurements using both LTE and HSPA technologies. Overall, about 2 hours of cellular traces have been collected. Fig.~\ref{fig_trace_subway_cross} and Fig.~\ref{fig_trace_times} show two samples of our traces. These samples clearly show the highly dynamic nature of cellular networks in which available link capacities vary fast (Traces are available at: \url{https://github.com/Soheil-ab/Cellular-Traces-2018}). 
In addition to our traces, we use the older data collected in prior work (\cite{mahi} and \cite{sprout}) from 5 different commercial cellular networks in Boston (T-Mobile's LTE and 3G UMTS, AT\&T's LTE, and Verizon's LTE and 3G 1xEV-DO).
\subsection{Schemes Compared and Metrics} 
In this section, we compare C2TCP with various schemes. We choose these schemes to cover different solution categories in our evaluation. In particular, we compare C2TCP with the state-of-the-art e2e schemes including Google's BBR \cite{bbr} (a delay and throughput based design), PCC-Vivace~\cite{vivace} (a delay-based online-learning equipped design), Verus~\cite{verus} (a delay-based TCP targeting cellular network), Sprout~\cite{sprout} (a delay-based design targeting cellular network), and different TCP flavors including Cubic~\cite{cubic} (the dominant and the most popular design on Internet) and Westwood~\cite{west} (an older scheme targeting cellular networks). We use 4 main performance metrics in this section: average throughput (in short, throughput), average and 95th percentile queuing delay, and jitter.

\begin{figure}[!t]
\centering
\includegraphics[width=0.5\textwidth,height=0.8in]{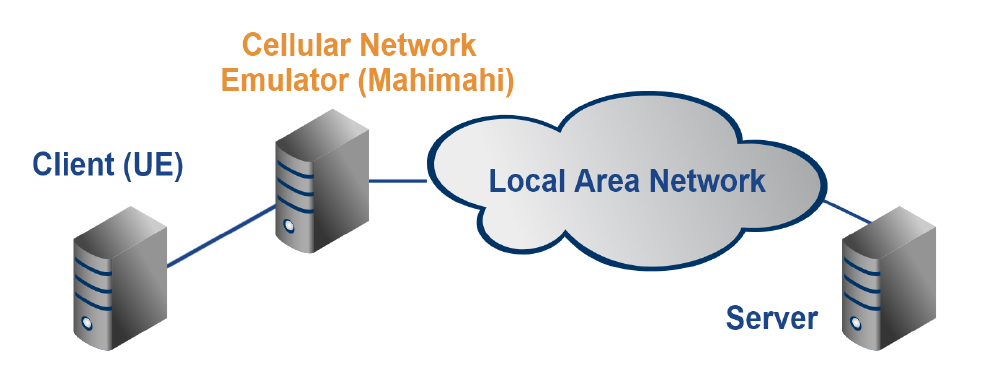}
\caption{Topology used for trace-driven evaluations}\label{fig_topo}
\end{figure}

\subsection{Topology} \label{sec_topology}We mainly use 3 entities (equipped with Linux OS) shown in Fig.~\ref{fig_topo} for these evaluations. The first one represents a server, the 2nd one emulates a cellular access channel using Mahimahi toolkit, and the 3rd one represents a UE. Similar to our traces, the minimum RTT is 20ms. Although the specific buffer size at BS for each client is not in public domain, we have tried to select the buffer size in our evaluations by comparing results from emulations with results from the real-world for a specific scheme such as Cubic. Based on that, the buffer size at bottleneck link is selected to be 150KB. Later, in section \ref{sec_queue}, we investigate the impact of the buffer size on the performance of C2TCP. For C2TCP, unless it is mentioned, we set Target to 50ms.

\subsection{Results}
\begin{figure*}[t!]
\centering
\includegraphics[width=\linewidth,height=3.7in]{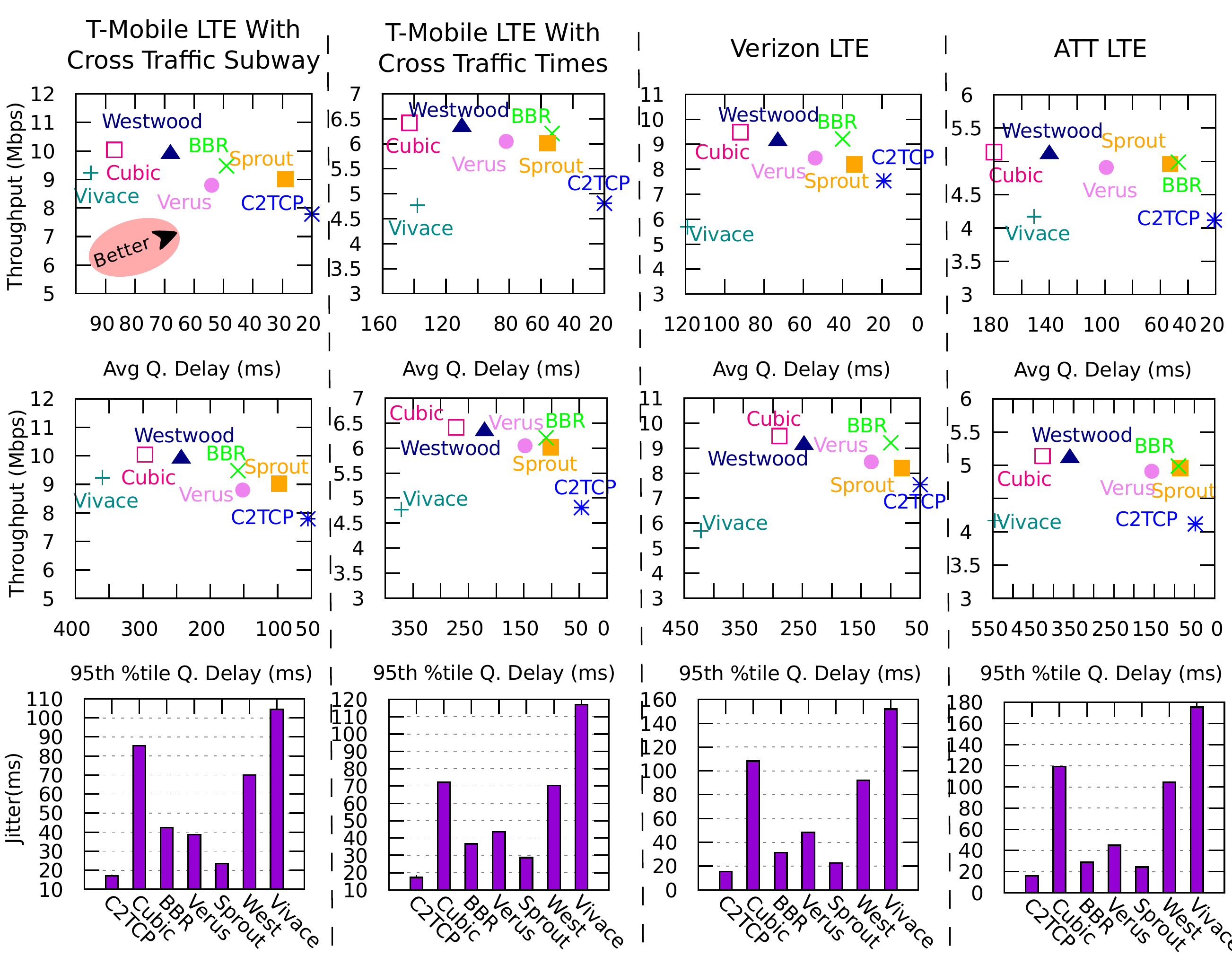}
\caption{Throughput, average queuing delay, Jitter, and 95th percentile queuing delay of each scheme over LTE cellular links}\label{fig_overall1}
\end{figure*}

\begin{figure}[!t]
\centering
    \includegraphics[width=\linewidth,height=3.4in]{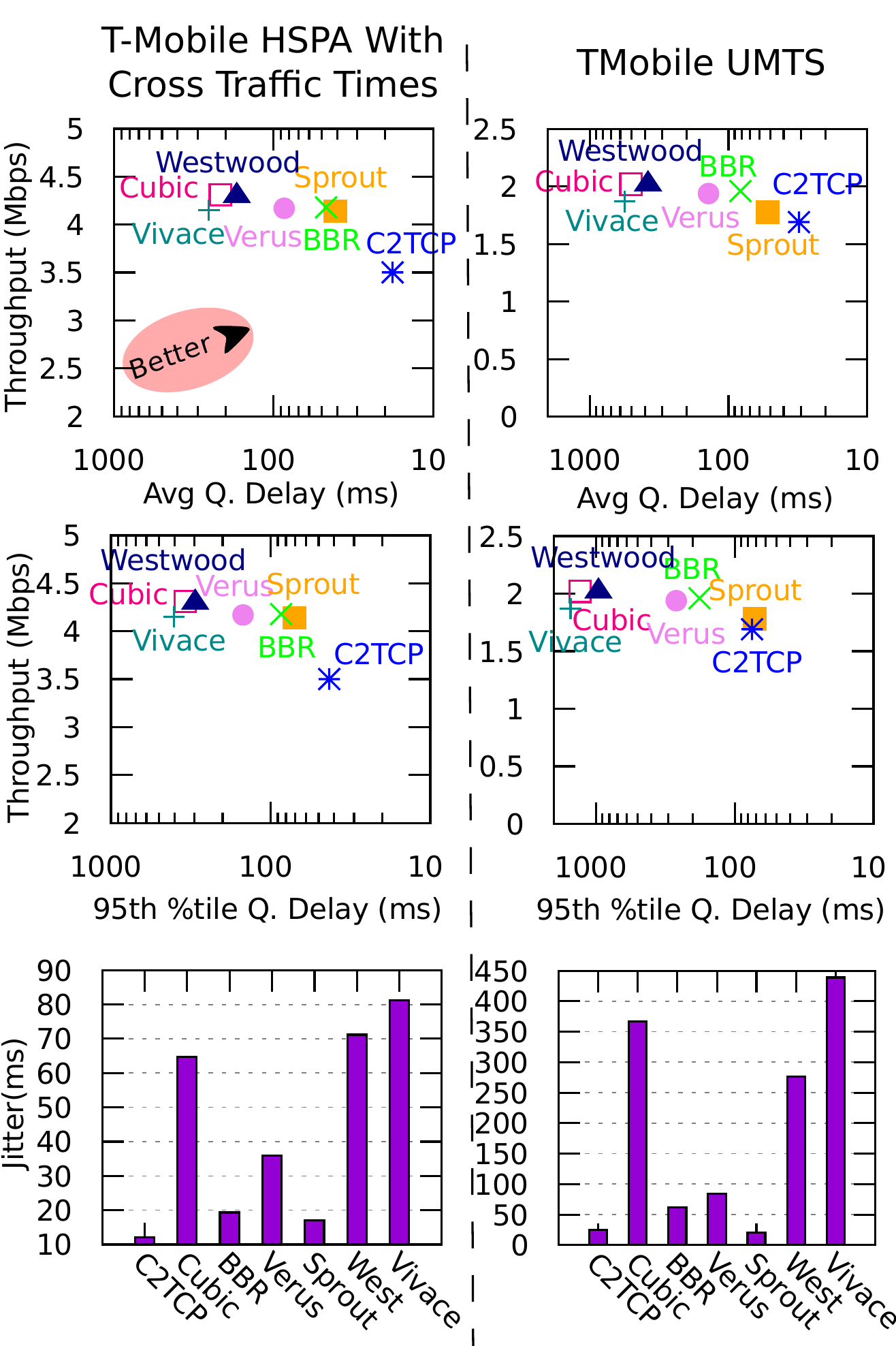}
    \caption{Throughput, average queuing delay, Jitter, and 95th percentile queuing delay of each scheme over UMTS and HSPA cellular links}\label{fig_overall2}
\end{figure}

Fig. \ref{fig_overall1} and Fig. \ref{fig_overall2} show the performance of various schemes in our extensive trace-driven evaluations for different traces. Due to space limitation, we only show the graphs for six traces. Results for other traces are similar to the ones shown here. In particular, Fig. ~\ref{fig_overall1} depicts results for LTE traces and Fig. ~\ref{fig_overall2} illustrates results  for UMTS and HSPA traces. For each trace, there are 3 graphs, one showing the average delay and throughput, one illustrating 95th percentile delay and throughput, and the other one showing the jitter performance. Schemes achieving higher throughput and lower delay (up and to right region of graphs) are more desirable. 

\begin{figure}[!t]
\centering
    \includegraphics[width=\linewidth,height=1.1in]{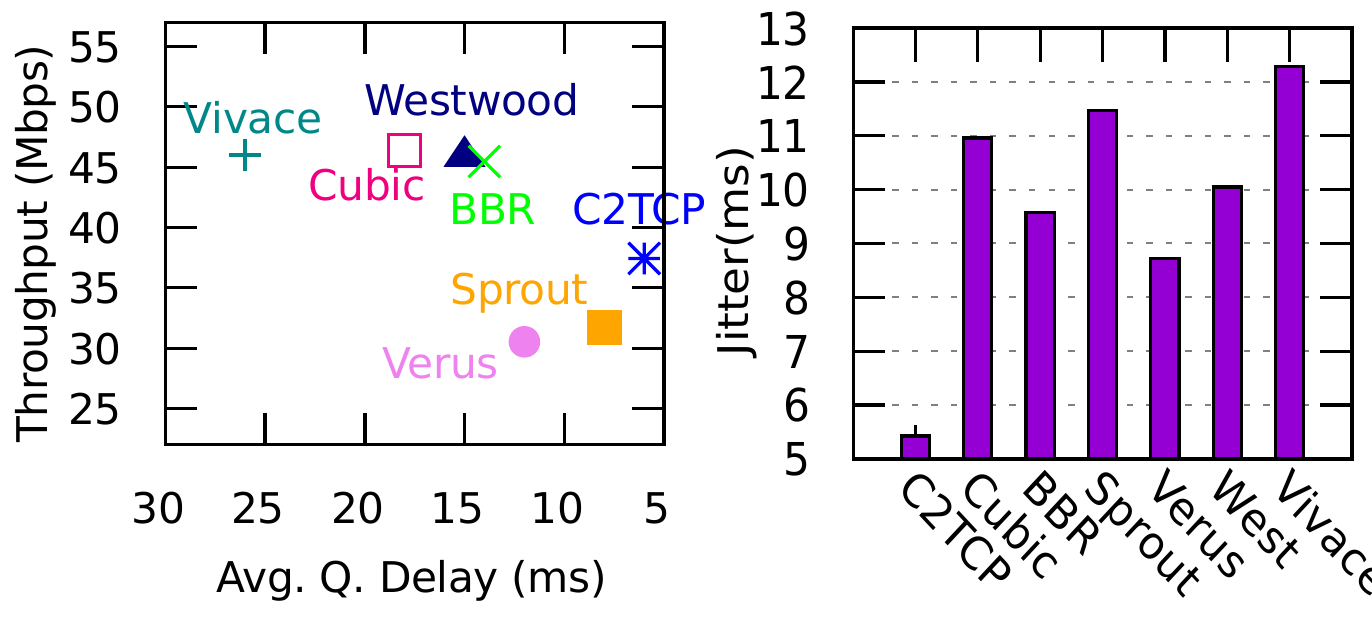}
    \caption{Throughput, average queuing delay, and jitter of each scheme for the Mobile Edge scenario}\label{fig_edge}
\end{figure}
\begin{table}[!t] \renewcommand{\arraystretch}{1} \caption{Overall normalized results averaged across all traces} 
\label{table_overall} \centering
\begin{tabular}{c|cccc}
\hline
& Throughput & Avg. Delay & Jitter & 95th\%tile Delay \\
\hline
C2TCP& 1 & \cellcolor{LightBlue}1 & \cellcolor{LightBlue}1 & \cellcolor{LightBlue}1 \\
BBR& 1.22 & 2.44 &  2.15 & 2.31\\
Verus& 1.12 & 3.82 & 9.25 & 3.22 \\
Cubic& \cellcolor{LightRed}1.28 & 8.95 & 7.19 & 8.54 \\
Sprout& 1.13 & 1.94 & 1.32 & 1.64 \\
Westwood& 1.26 & 6.89 & 6.04 & 6.78 \\
PCC-Vivace& 1.04 & 10.05 &  9.25 & 10.52\\
\hline
\end{tabular}
\end{table}
\begin{figure*}[t!]
    \begin{minipage}[b]{\linewidth}
        \centering
        \includegraphics[width=\textwidth,height=2.2in]{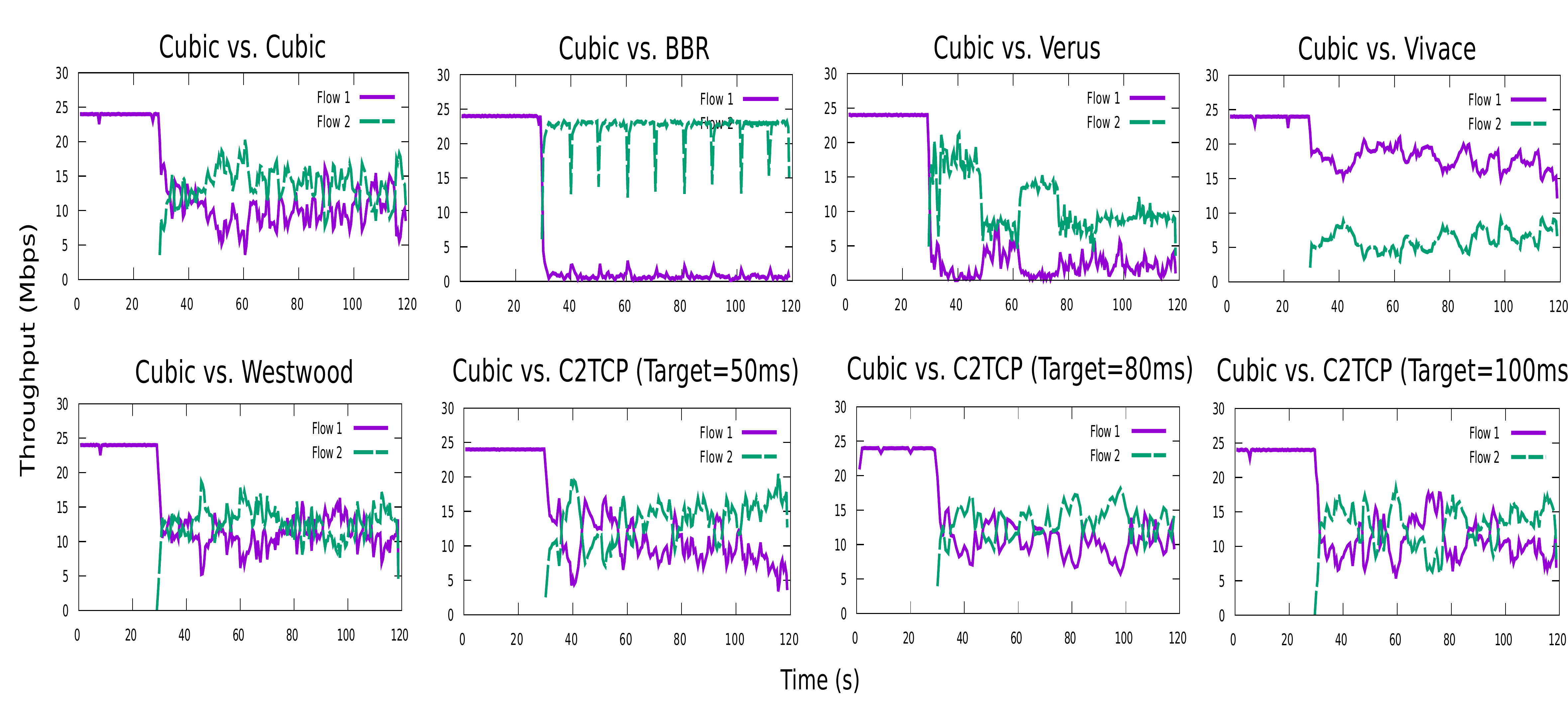}
        \caption{Share of bandwidth among Cubic, started at time $0$ (Flow 1), and other schemes, started at $30s$ (Flow 2with dashed lines on each graph)}\label{fig_fair}
    \end{minipage}
\end{figure*}
We have normalized results of different schemes for various traces to C2TCP's performance and averaged them through all evaluations. The overall averaged normalized results through all traces are shown in Table~\ref{table_overall}\footnote{It is worth mentioning that all experiments have been repeated several times to make sure that the results presented here are not affected by the random variations.}. C2TCP achieves the lowest average delay, the lowest jitter, and the lowest 95th percentile delay among all schemes, while compromising throughput slightly. For instance, compared to Cubic, C2TCP decreases the average delay by about $9\times$, while compared to Cubic which achieves the highest throughput, it only compromises throughput by $0.28\times$.

Generally, results for different traces in Fig.~\ref{fig_overall1} and Fig.~\ref{fig_overall2} show a common pattern. As expected, Cubic achieves the highest throughput among all schemes. The reason is that since it is not sensitive to delay, it simply builds up the queue. Therefore, it will achieve higher utilization of the cellular access link when the channel experiences good quality. In that sense, Westwood is similar to Cubic, though it performs slightly better than Cubic with a smaller delay. Verus performs better than schemes such as Cubic and achieves lower average and 95th percentile delays. However, its delay performance is far from the delay performance of Sprout, BBR, and C2TCP for almost all traces. Design of BBR is based on first getting good throughput and then reaching good delays~\cite{bbr}. BBR tries to find the bottleneck link bandwidth by periodically increasing its sending rates. Although that might work in a wired network where the bottleneck link bandwidth does not change very fast, in a highly dynamic environment such as a cellular network, it fails. Therefore, there are times that BBR sends packets with the rates higher than bottleneck link bandwidth. Therefore, as Fig.~\ref{fig_overall1} and Fig.~\ref{fig_overall2} show it experiences queuing delay. The main idea behind Sprout is to predict the future of cellular link's capacity and send packets to the network cautiously to achieve low 95th percentile delay. Although Sprout can achieve good delay performance, C2TCP still beats its delay performance by about $2\times$. PCC-Vivace, as admitted by its authors, cannot perform well in a highly dynamic network such as cellular networks. The main reason is a relatively long time that it requires to converge to its targeted rate. 
C2TCP controls the average delay and keeps it below the application's Target while having a high throughput. Results confirm that C2TCP performs well across all traces while maintaining a good throughput performance. For the other technologies such as HSPA and UMTS where the cellular network intrinsically experiences lower throughput and higher delay, C2TCP still outperforms other schemes (Fig.~\ref{fig_overall2}) and achieves a smaller controllable delay and good throughput. 

\subsection{Mobile Edge Scenario}
Now, we examine C2TCP's capability to achieve very low average delays in the mobile edge computing architecture where the server application is close to base station at the edge and show that C2TCP can achieve very low delays such as 10ms. To that end, we set the minimum RTT of the topology shown in Fig.~\ref{fig_topo} to 4ms. To make it feasible to achieve very low average delays in the network, we used a modified version of the cellular trace shown in Fig.~\ref{fig_trace_times}. For the modified trace, we increased the link capacity 3$\times$ at each arbitrary time to have a long-term average line capacity of 46Mbps. We set Target value to 10ms and compare C2TCP's average queuing delay, throughput, and jitter performance with other schemes. Fig.~\ref{fig_edge} shows the results. C2TCP achieves stringent Target delay of 10ms while having the lowest jitter among all schemes (nearly $2\times$ less jitter than the second best performing scheme) while compromising throughput only at most 20\% (compared to the best throughput achieved by Cubic). 
\section{C2TCP Micro-Evaluation}
\label{sec_eval_micro}
In this section, we look into more characteristics of C2TCP. In particular, we investigate C2TCP's friendliness to existing TCP flows (e.g., Cubic), its fairness to other C2TCP flows, the impact of changing Target on its performance, comparison of our e2e solution with CoDel, an in-network AQM design, and impact of the buffer size on the performance of C2TCP.

\subsection{TCP Friendliness}
\label{sec_fair}
Before reaching the last mile (BS to UE), C2TCP flows will need to go from the server to BS. Therefore, it will most likely coexist with other TCP flows in network's switches. So, in this section, we investigate an important requirement of any TCP variant: TCP Friendliness. TCP friendliness property means that in the presence of other TCP variants, how fair the bandwidth will be shared among the competing flows. Usually, a scheme that is too aggressive is not a good candidate since it may starve flows of other TCP variants. 

To evaluate the C2TCP's TCP friendliness, we use Mahimahi~\cite{mahi} to connect two servers to one client using a normal switch. In particular, we send one Cubic flow from one server to the client. Choosing Cubic as the reference TCP rests on the fact that Cubic is the default TCP in Linux and Android OS which takes more than 60\% of smartphone/tablet market share\cite{android-market}. Then, after 30 seconds, we start sending another flow from the second server to the client using different schemes including Cubic, BBR, Verus, PCC-Vivace, Westwood, and C2TCP. When there is a very large queue in the switch, there will be no scheme which can get a fair portion of bandwidth when the queue is already being filled up by another aggressive flow~\cite{sprout}. So, to have a fair comparison, as a rule of thumb, we set the buffer size of the switch to the BDP (bandwidth delay product) of the network. Here, the access link's bandwidth, RTT, and the buffer size are 24Mbps, 20ms, 40 packets (1pkt=1.5KB), respectively. Also, we use different Target delays for C2TCP to examine the impact of it on friendliness property of C2TCP. In particular, we set Target to 50ms, 80ms, and 100ms. \footnote{Sprout's~\cite{sprout} main design idea is to model the cellular access link bandwidth using a varying Poisson process, so this scheme won't work properly when link bandwidth is constant. Therefore, to have a fair comparison, we don't include performance results of this scheme here.} 

Fig.~\ref{fig_fair} shows the average throughput gained by different schemes throughout time. The results indicate that BBR and Verus are aggressive and will get nearly all the bandwidth from the Cubic flow, while PCC-Vivace's share of link's bandwidth cannot grow in the presence of Cubic. 

The main idea of BBR is to set congestion window to the BDP of the network. To do that, it measures min RTT and the delivery rate of the packets. When the buffer size is at the order of BDP, BBR fully utilizes the queue and will not leave room for other flows. Therefore, here when BBR shares the queue with another Cubic flow, the Cubic flow experiences extensive packet drops and won't achieve its fair share of the bandwidth. PCC-Vivace changes the sending rate and tracks its impact on a predefined utility function. However, the fact that it requires time to figure out the good rates make it suffer from the presence of Cubic flow in the queue. In both cases, either being very aggressive (BBR) or being too moderate (PCC-Vivace), the TCP friendliness characteristic of these schemes is not desirable. 

\begin{figure}[!t]
\centering
    \includegraphics[width=0.5\textwidth,height=1.1in]{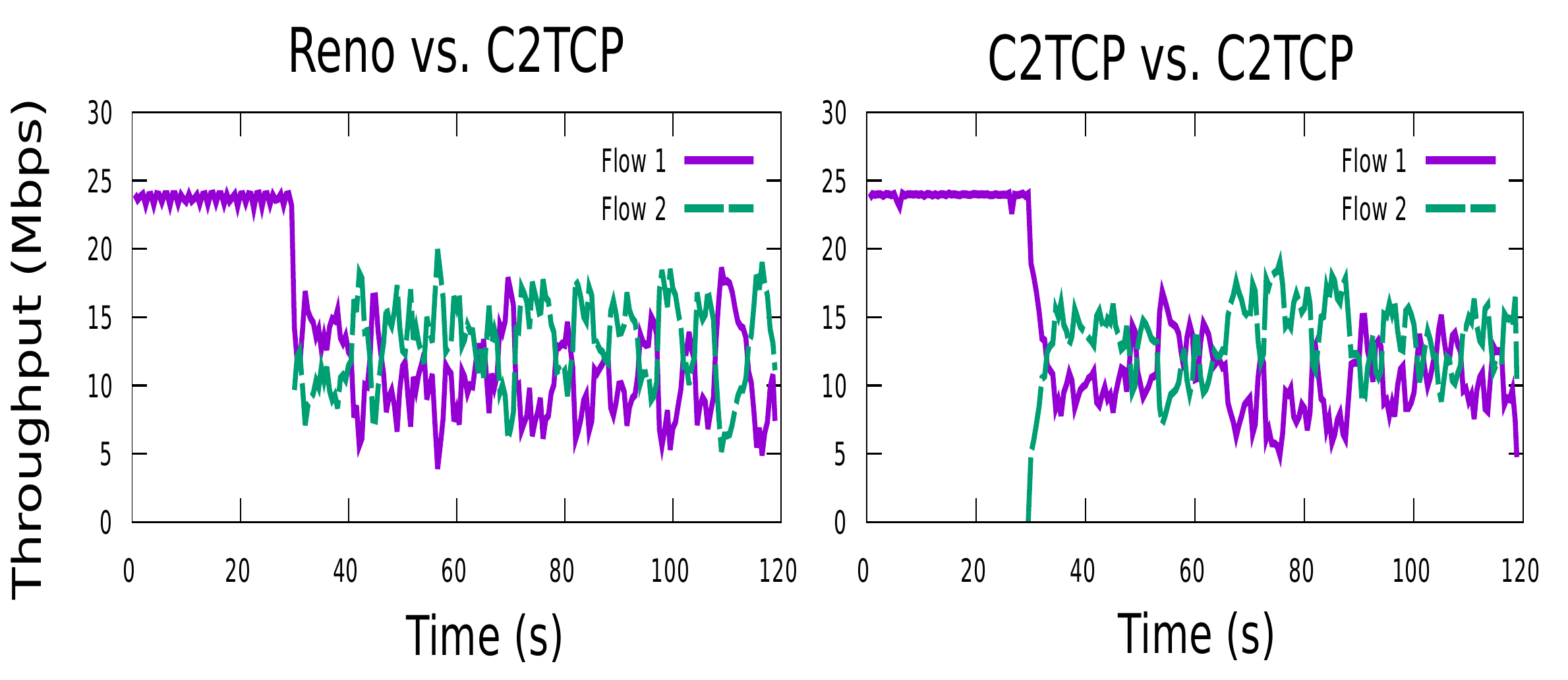}
    \caption{Share of bandwidth among a NewReno flow (started at 0) and a C2TCP flow (the left graph), and two C2TCP flows (the right graph)}\label{fig_fair2}
\end{figure}

However, as Fig.~\ref{fig_fair} illustrates, flows of Westwood, Cubic, and C2TCP  can share the bandwidth with the Cubic flow fairly. C2TCP is implemented over Cubic. So, to show that C2TCP's fairness property is not because the competing flow in the test is Cubic, we replace Cubic flow with a NewReno flow and do the test again. The result is shown in Fig.~\ref{fig_fair2} (the left one). Also, we evaluate the fairness between two C2TCP flows (both C2TCP flows have 80ms Target) and the result is shown in Fig.~\ref{fig_fair2} (the right one). Fig. \ref{fig_fair2} (the right graph) depicts that C2TCP is fair to the other C2TCP flow in the network. That being mentioned, C2TCP is friendly to other TCP flows and can achieve good fairness property with other C2TCP flows. 
\subsection{Impact of Target and Comparison with an In-Network Scheme}
\label{sec_target}
\label{eval-codel}
In this section, we change the application's Target of average delay and investigate the impact of it on the overall performance of C2TCP. Also, we compare the performance of C2TCP, an end-host solution, with CoDel, an in-network solution which is one of the schemes that inspired us. To do that, we add CoDel AQM algorithm to both uplink and downlink queues in Mahimahi and use Cubic at the end hosts. Here, a cellular trace shown in Fig.~\ref{fig_trace_times} has been used for the experiments.  

\begin{figure}[!t]
\centering
    \centering
    \includegraphics[width=0.8\linewidth,height=1.3in]{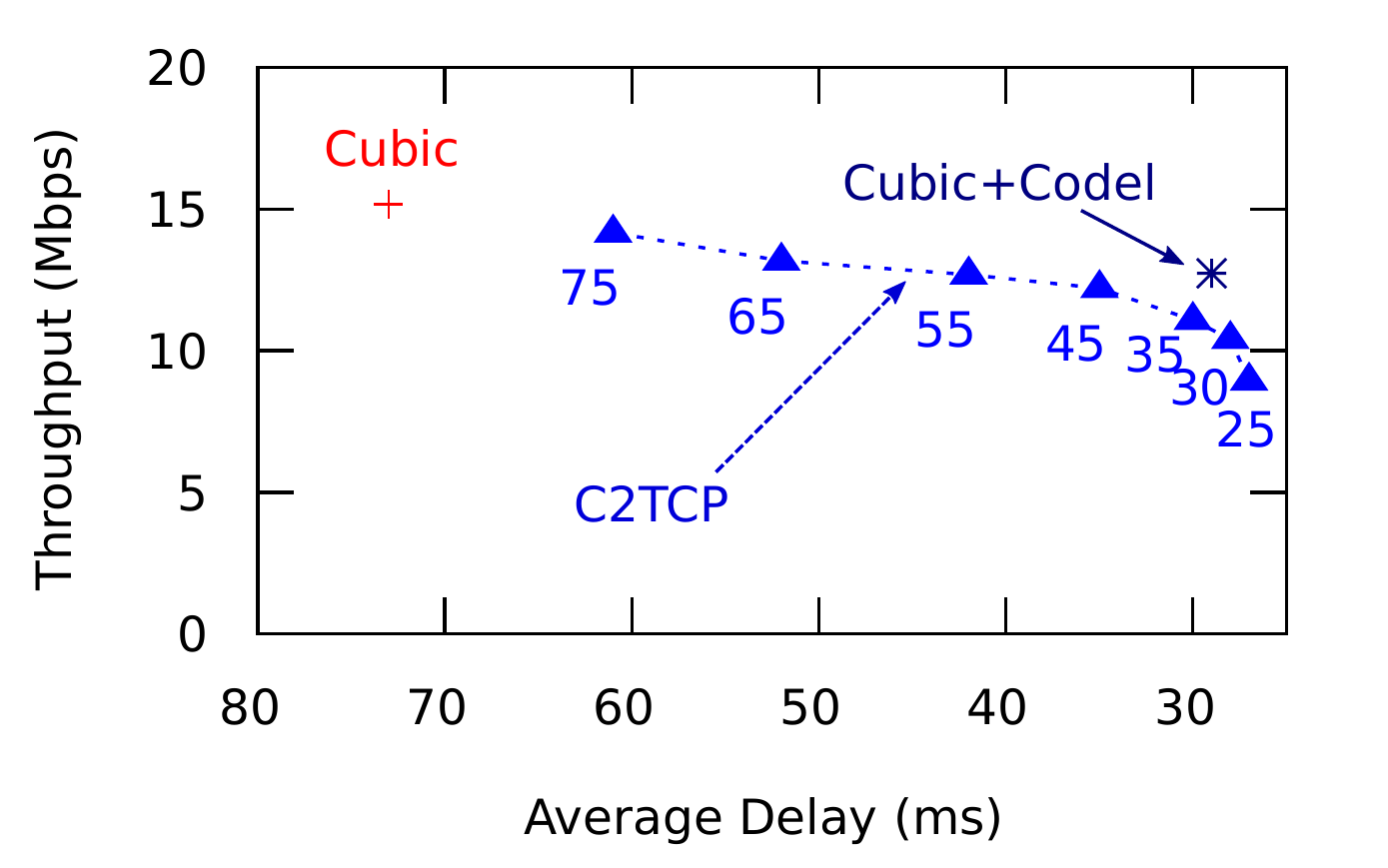}
    \caption{Impact of Target Values on Throughput and Delay (numbers on the graph show the chosen Target of the application in the experiment)}\label{fig_target}
\end{figure}

In particular, we vary the Target value from 25ms to 75ms. The average e2e delay (average queuing delay plus the minimum RTT of the network, i.e., 20ms) and throughput achieved for each setting are shown in Fig.~\ref{fig_target}. By varying the value of Target, an application can control its average packet delay while achieving a good throughput. As expected, increasing Target will push C2TCP toward Cubic's performance. 

\begin{figure}[!t]
    \centering
        \includegraphics[width=\linewidth,height=1.3in]{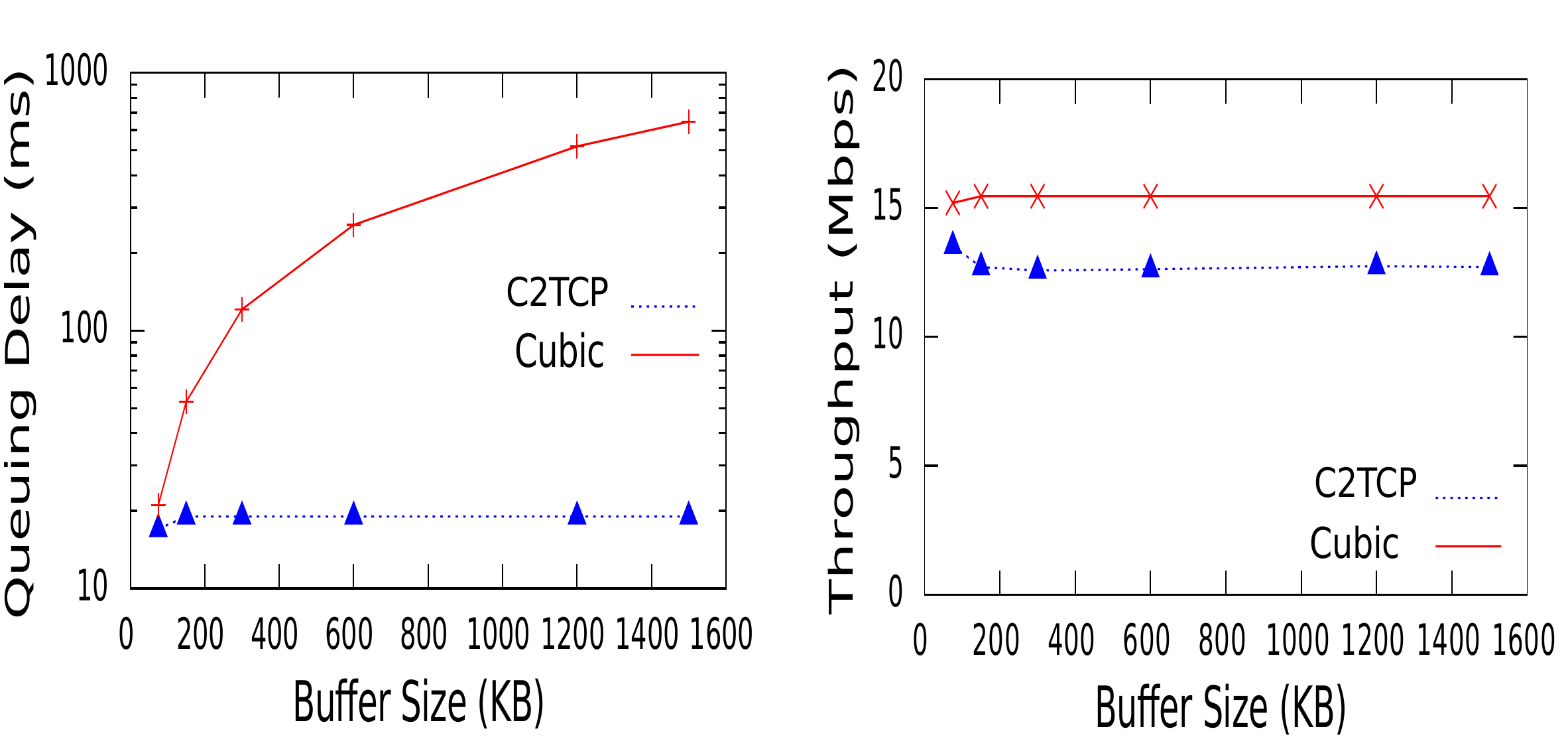}
        \caption{Impact of buffer size on performance of C2TCP and Cubic}\label{fig_queue}    \end{figure}

Using CoDel improves the delay performance of Cubic while degrading its throughput. As Fig.~\ref{fig_target} illustrates, C2TCP can achieve even better delay performance than CoDel when application's Target is chosen accordingly at the cost of compromising throughput. It is worth mentioning that to have in-network solutions such as CoDel, cellular carriers must install them inside their base stations and in base band modem or radio-interface drivers on cellular phones, while an end-host solution scheme like C2TCP only requires to update the software at the server, and thus is much more deployment friendly than in-network solutions. 

Also, C2TCP provides great flexibility for applications when it is compared to in-network schemes such as RED and CoDel in which a set of queue parameters are set for all applications. From the aspect of the architecture design, it is more desirable to have an end-host solution providing a degree of freedom for applications to control their operational points instead of having an in-network solution lacking the flexibility of accommodating various application requirements and requiring modified gateways and network devices.

\subsection{Impact of Buffer Size on C2TCP's Performance}
\label{sec_queue}
Deep per-user buffers at BS is one of the cellular network characteristics that distinguish them from the wired network in which all users will compete for the same queue which is not usually very big. One of the reasons for such a design is that cellular network providers try to increase their network's reliability and drop packets as few as possible. However, this leads to the well-known problem of bufferbloat~\cite{bufferbloat}. 

C2TCP's main design goal is to control the delay of packets. So, one of its primary features and important properties is that it has very low sensitivity to the size of the queue. To show that, here, we explore the impact of the buffer size on C2TCP's performance. In particular, we use one of our traces shown in Fig.~\ref{fig_trace_times} and vary the bottleneck buffer size from 75KB to 1.5MB and compare the queuing delay and throughput of C2TCP and its parent loss-based TCP - Cubic. Results are shown in Fig.\ref{fig_queue}. As expected, Cubic's delay performance degrades dramatically when buffer size is increased. In contrast, C2TCP's delay performance is independent of the buffer size, though it actually built on top of Cubic. On the other hand, both schemes achieve almost constant throughput performance, though for different reasons. Cubic always occupies the bottleneck link queue as much as it can to utilize the bottleneck link bandwidth. In contrast, C2TCP tries to always send a proper amount of packets (which is almost independent of the buffer size) into the network to control the delay of them. So, change in the bottleneck buffer size will not affect its throughput performance as shown in Fig.~\ref{fig_queue}.
 

\section{Discussion}
\textbf{Does C2TCP work in other networks?} Our design rests on the underlying architecture of cellular networks including the presence of deep per-user buffers at BS, exploiting a scheduler at BS which brings fairness among various UEs at the bottleneck link (last-mile), and low e2e feedback control delay (thanks to current technologies and trends such as MEC, MCDN, M-CORD~\cite{mcord}, etc.). Therefore, the lack of this kind of structure will affect C2TCP's performance. For instance, for networks with very large intrinsic RTTs, end-hosts absorb the network's condition with a large delay due to the large feedback delay. Therefore, because of that large feedback delay, C2TCP (and any other e2e approaches) couldn't catch fast link fluctuations and respond to them fast enough. 

\textbf{Abusing the parameters:} Misusing a layer 4 solution and setting its parameters to get more share of the network bandwidth by users is always a concern. For instance, a user can change the initial congestion window of loss-based schemes such as Cubic in Linux kernel. Similarly, users might be tempted to abuse the Target parameter of C2TCP. Although providing mechanisms to prevent these misuses is beyond the scope of this paper, our minimum and maximum values for $\alpha$ can mitigate the issue. In addition, in TCP, the sender's congestion window will be always capped to the receiver's advertised window (RcvWnd). 

\textbf{C2TCP flows with different requirements for a user:}
When a cellular phone user runs a delay sensitive application (such as real-time gaming, video conferencing, virtual reality content streaming, etc.), flows of that application are the main interested flows (highest priority ones) for the user. Therefore, throughout the paper, we have assumed that it's rare to have flows of other applications with different delay requirements competing with the highest priority flows for the same user. However, in case of having multiple applications with different requirements for the same user, we think that any transport control solution (such as Cubic, Sprout, C2TCP, etc.) should be accompanied with prioritization techniques at lower layers to get good results in practice (e.g. \cite{hyline,ffq}). For instance, one simple solution is to use the strict priority tagging for packets of different flows (by setting differentiated services field in the IP header) and serve flows based on these strict priorities in the network.

\textbf{Setting Target:} 
Instead of setting Target value per application, we could set it per class of applications. In other words, we could let applications choose their application types. Then, C2TCP would set the Target using a table including application types and their corresponding Target values configured in an offline manner. 

\section{Related Work}
\textbf{e2e congestion control protocols:}
Congestion control is always one of the hottest topics with huge studies including numerous variants of TCP. TCP Reno\cite{reno}, TCP Taho\cite{tahoa}, and TCP NewReno\cite{newreno} were among early approaches using loss-based structures to control the congestion window. TCP Vegas\cite{vegas} tries to do congestion control directly by using measured RTTs. TCP Cubic\cite{cubic} changes incremental function of the general AIMD-based congestion window structure, and Compound TCP \cite{compound} maintains two separate congestion windows for calculating its sending rate. PCC-Vivace \cite{vivace} uses online learning techniques to choose best sending rates. BBR \cite{bbr} estimates both maximum bottleneck bandwidth and minimum RTT delay of the network and tries to work around this operation point, though \cite{jaf} has proved that no distributed algorithm can converge to that operation point. Also, LEDBAT\cite{ledbat}, BIC\cite{bic}, and TCP Nice\cite{nice} are among other TCP variants. However, all these schemes are mainly designed for a wired network, i.e., fixed link capacities in the network. In that sense, they are not suitable for cellular networks where link capacity changes dynamically. 

Among the state-of-the-art proposed schemes targeting cellular networks, Sprout~\cite{sprout} and Verus ~\cite{verus} are worth being mentioned. Sprout introduces a stochastic forecast framework for predicting the bandwidth of the cellular link, while Verus tries to make a delay profile of the network and then use it to calculate congestion window based on the current network delay. TCP Westwood \cite{west}, which is among the older designs targeting wireless networks, attempts to select a slow start threshold more consistent with the actual available bandwidth and introduces a new fast recovery scheme. We have compared C2TCP with most of these schemes in section~\ref{eval}.

\textbf{AQM schemes and feedback-based algorithms:}
Active queue management schemes (such as RED~\cite{red}, BLUE \cite{blue}, and AVQ \cite{avq}) use the idea of dropping/marking packets at the bottleneck links so that end-points can react to packet losses and control their sending rates. It is already known that automatically tuning parameters of these schemes in the network is very difficult~\cite{codel,sprout}. To solve this issue, CoDel~\cite{codel} proposes using sojourn time of packets in a queue instead of queue length to drop packets to indirectly signal the end-points. However, even this improved AQM scheme still has a profound issue inherited from its legacy ones: these schemes all seek a ``one-setting-fits-all'' solution, while different applications might have different throughput or delay requirements. Even one application can have different delay/throughput requirements during different periods of its lifetime. 

Also, there are different schemes using feedback from the network to do a better control over congestion window. Among them, various schemes using ECN\cite{ecn} as the main feedback. The most recent example is DCTCP\cite{dctcp} which changes congestion window smoothly using ECN feedback in datacenter networks. However, DCTCP  similar to other TCP variants is mainly designed for stable link capacities but not highly variable cellular link capacities.

AQM and feedback-based schemes have a common problem: they need changes in the network which is not desirable for cellular network providers due to high CAPEX costs. Inspired by AQM designs such as CoDel and RED, C2TCP provides an e2e solution to circumvent the problem. Our approach does not require any change/modification/feedback to/from the network.


\section{Conclusion}
We have presented C2TCP, a congestion control protocol designed for cellular networks to control the delay of packets and achieve high throughput. Our main design philosophy is that achieving good performance does not necessarily comes from complex rate calculation algorithms or complicated channel modelings. C2TCP works on top of classic throughput-oriented TCP and provides it with a sense of delay without using any network state profiling, channel prediction, or complicated rate adjustments mechanisms. This enables C2TCP to achieve ultra-low latency communications for the next generation of highly delay-sensitive applications such as AR/VR without the need for changing network devices (It only modifies the server side). Our real-world experiments and trace-driven evaluations show that C2TCP outperforms well-known TCP variants and existing state-of-the-art schemes which use channel prediction or delay profiling of the network.

\section*{Acknowledgment}

We would like to thank Siwei Wang for his help on collecting the trace files and Tong Li for his help on simulations of the earlier version of this work. We also would like to thank anonymous JSAC reviewers whose comments helped us improve the paper.

\bibliographystyle{IEEEtran}

\ifCLASSOPTIONcaptionsoff
  \newpage
\fi

\end{document}